\documentclass{jfm}
\usepackage{graphicx}
\usepackage{epstopdf, epsfig}

\usepackage{amsmath}
\usepackage{nccmath}
\usepackage{xcolor}
\usepackage{subcaption}

\newcommand{\nvn}[1]{\textcolor{purple}{#1}}

\usepackage[normalem]{ulem}
\usepackage{nicefrac}

\shorttitle{Dense Suspensions in Rotary Shear}
\shortauthor{N. K. Agrawal, Z. Ge, M. Trulsson, O. Tammisola and L. Brandt}

\title{Dense Particle Suspensions in Rotary Shear Flows}

\author{Naveen Kumar Agrawal\aff{1},
  Zhouyang Ge\aff{1,2},
  Martin Trulsson\aff{3}
\corresp{\email{martin.trulsson@compchem.lu.se}},
  Outi Tammisola\aff{1}
  \and Luca Brandt\aff{1,4,5}
  }

\affiliation{\aff{1}FLOW Centre and Department of Engineering Mechanics, KTH Royal Institute of Technology, SE-100 44 Stockholm, Sweden

\aff{2}Department of Mechanical Engineering and Institute of Applied Mathematics, University of British Columbia, Vancouver V6T 1Z4, British Columbia, Canada

\aff{3} Computational Chemistry, Lund University, Lund SE-221 00, Sweden

\aff{4}Department of Environment, Land and Infrastructure Engineering (DIATI), Politecnico di Torino, 10129 Turin, Italy

\aff{5} Department of Energy and Process Engineering, Norwegian University of Science and Technology (NTNU), Trondheim, Norway
}

\date{\today}

\begin{document}

\maketitle

\begin{abstract}
We introduce a novel unsteady shear protocol, which we name Rotary Shear (RS), where the flow and vorticity directions are continuously rotated around the velocity gradient direction by imposing two out-of-phase oscillatory shear (OS) in orthogonal directions. We perform numerical simulations of dense suspensions of rigid non-Brownian spherical particles at volume fractions ($\phi$) between 0.40 and 0.55 subject to this new RS protocol and 
compare to the classical OS protocol. We find that the suspension viscosity displays a similar non-monotonic response as the strain amplitude ($\gamma_0$) is increased: a minimum viscosity is found at an intermediate, volume-fraction dependent strain amplitude. However, the suspension dynamics is different in the new protocol. Unlike the OS protocol, suspensions under RS do not show self-adsorbing states at any $\gamma_0$ and do not undergo the reversible-irreversible transition:
the stroboscropic particle dynamics are always diffusive, which we attribute to the fact that the RS protocol is irreversible. To validate this hypothesis, we introduce a reversible-RS (RRS) protocol, a combination of RS and OS, where we rotate the shear direction (as in RS) until it is instantaneously reversed (as in OS), and find the resulting rheology and dynamics to be closer to OS. Detailed microstructure analysis shows that both the OS and RRS protocols result in a contact-free, isotropic to an in-contact, anisotropic microstructure at the dynamically reversible-to-irreversible transition. The RS protocol does not render such a transition, and the dynamics remain diffusive with an in-contact, anisotropic microstructure for all strain amplitudes.
\end{abstract}

\section{Introduction}\label{sec:introduction}
Suspensions are one of the simplest complex fluids to make, consisting solely of solid particles suspended in a viscous fluid. Despite the simplicity, driven suspensions can show a multitude of fascinating and complex physics \citep{guazzelli2018rheology, ness2022physics}. For example, under deformation such as shear, they can jam, thicken, or thin, depending on the particle-particle interactions and the solid content, \emph{i.e.}, volume fraction \citep{seto2019shear}. Understanding when and how the suspension mechanical properties, such as viscosity and normal stress differences, change upon changing conditions is crucial for an effective processing of these materials, not the least as they are vividly used in industry (\emph{e.g.,} in food, cosmetic, personal care, and drug processing). 

For dense suspensions composed of rigid particles under excluded-volume interactions, with or without frictional forces, the steady-state viscosity is only a function of the (solid) volume fraction and shows a power-law divergence around the shear jamming fraction \citep{Zarraga2000characterization, ovarlez2006local, olsson2007critical, bonnoit2010inclined, andreotti2012shear, trulsson2012transition}. 
However, in the case of unsteady flow conditions, for example in oscillatory shear (OS) with only excluded-volume interactions, the viscosity is found to depend not just on the packing fraction but also on the strain amplitude \citep{breedveld2001shear, bricker2006oscillatory, bricker2007correlation, lin2013short, ness2017oscillatory, martone2020non, ge2022rheology}. The strain amplitude dependence shows a non-monotonic behavior of the viscosity, with minima at an intermediate strain amplitude which depends on the volume fraction.

The particle dynamics also depend on the strain amplitude in OS. 
In $1966$, G.~I.~Taylor illustrated the reversibility feature of Stokes flow in a now-classic experiment: By reversing the direction of the flow, the dye particles suspended in a viscous Newtonian fluid returned to their initial positions. However, later studies have shown that neutrally buoyant non-Brownian particles in fact \emph{diffuse} in a sheared Stokesian fluid \citep{eckstein1977self,  leighton1987measurement,davis1996hydrodynamic, marchioro2001shear,sierou2004shear}. 
This apparent contradiction was reconciled by the chaotic nature of hydrodynamic interactions between the particles because there is inevitably noise in the system, leading to the loss of memory and correlation \citep{drazer2002}.
More recently, \citet{pine2005chaos} and others \citep{corte2008random, menon2009universality, metzger2010irreversibility, metzger2013irreversibility, Ness_Cates_PRL2020, ge2021irreversibility, mari2022absorbing}, demonstrated that non-Brownian suspensions subject to periodic shear flow can undergo a dynamical phase transition, called ``reversible-irreversible transition" (RIT). 
Remarkably, when the amplitude is low, particles self-organize at the end of each period to avoid further collisions. 
The phenomenon, termed random organization, leads to reversible particle trajectories in idealized situations and was first discovered for dilute suspensions, but recently also seen in the semi-dilute and the dense regimes. In many real cases, the dynamics do not become perfectly reversible but rather sub-diffusive, due to thermal or other noise sources in the system. Nevertheless, both perfect reversibility and sub-diffusivity are a signature of the onset of  random organization \citep{pine2005chaos}. 
On the other hand, above a critical strain amplitude, the particle motion becomes diffusive and irreversible, consistently with the chaotic dynamics observed in simple shear flows. 
Furthermore, it was found that this critical strain amplitude ($\gamma_{0,c}$) decreases with the volume fraction ($\phi$) as $\gamma_{0,c} \sim \phi^{-2}$ \citep{pine2005chaos}, suggesting that diffusion results from many-body interactions. 

The above studies clearly show that suspensions in OS display intricate rheological and dynamical behaviours, both of which depend on the volume fraction and driving strain amplitude.
As these properties are ultimately determined by the suspension microstructure (assuming no friction or time-dependent interactions), it is reasonable to expect a relation between the suspension rheology and dynamics.
Indeed, \cite{ge2022rheology} have recently shown that at the strain amplitude where the suspension viscosity is minimal, the suspension dynamics also transition from reversible to diffusive.
Furthermore, these authors found other rheological signatures, including an enhanced intracycle shear thinning and a finite second normal stress difference, at the onset of RIT.
These observations are intriguing and suggestive of a unique relationship between the dynamics and rheology; however, we note that they were made in periodically sheared suspensions where RIT \emph{would} occur. 
To determine whether there is indeed such a unique relationship requires exploring new flow protocols where the underlying dynamics have not been characterized.

Although most of the previous works have been devoted to oscillatory shear flow, with inherent periodicity and sudden shear reversal, the characterization of the mechanical response to changes in \emph{shear axes} has seldom been considered. The only exception is perhaps the recent efforts of \citet{blanc2023rheology} and \citet{acharya2024optimum}, who investigated the rheology (but not the dynamics) of dense suspensions after a \emph{sudden rotation} of the shear axes. These authors observed a drop in shear viscosity after shear rotation on strain scales of the order of unity or less. The magnitude of the viscosity drop was found to depend on the rotation angle, from no drop for 0-degree rotation, since this case corresponds to no change, to a maximum drop at 180-degree rotation around the gradient axis, corresponding to a classical shear reversal experiment as previously studied by \emph{e.g.,} \citep{peters2016rheology}, with a gradual and smooth transition between these two extremes. Alternating shear rotations also reduce the average viscosity and dissipation, if a small strain is chosen between repeated rotations \citep{acharya2024optimum}.
Despite the novelty of the shear rotation protocol, the change in the shear direction always occurs suddenly. This raises an interesting question: What if, instead,  the gradient shear axis is rotated \emph{smoothly} and \emph{gradually}?

In this work, we aim to fill this gap by investigating both the rheology and dynamics of dense suspensions beyond the sudden shear reversal and sudden shear rotation.
To this end,  we propose a novel shear protocol called rotary shear (RS), where we slowly change the shear direction in a continuous way. This new rotary shear flow, which is a linear combination of two out-of-phase oscillatory shears perpendicular to each other, has the benefit of not relying on any shear reversals (\emph{i.e.,} a sudden change in sign/direction),  has a constant shear magnitude, and it is still periodic. Hence, the microstructure might evolve in a gentler way.  We investigate this new rotary shear protocol in terms of both mechanical properties and microstructural changes, numerically comparing it to the extensively studied OS protocol to find the correspondence between the rheology and dynamics. Ultimately, our findings will help to link the rheological and dynamical behaviors of a dense suspension through its microstructure.

The outline of the paper is as follows: We start by discussing our numerical model in \S \ref{sec:model}. In \S \ref{sec:results} we show the RS's suspension rheology, its dynamics, and the microstructure in comparison with OS and try to understand the difference and similarity between the two protocols using a reversible combination of the two: RRS. We summarise our main findings in \S \ref{sec:conclusion}.

\section{Modelling and Method}\label{sec:model}
\begin{figure}
\centering
\includegraphics[width=\textwidth]{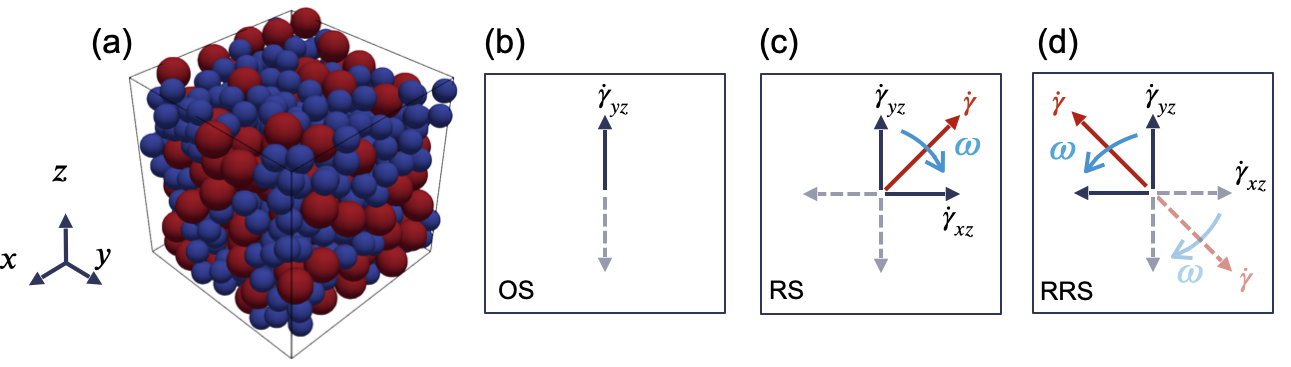}
\caption{(a) Schematic view of the dense suspension. Top view showing the directions of shear rate for (b) oscillatory shear (OS) (c) rotary shear (RS), and (d) reversible rotary shear (RRS), the dashed red line shows shear direction just before the reversal and the solid red line shows the instantaneous shear direction. Here, we denote the vorticity, streamwise, and velocity-gradient directions as $x$,$y$, and $z$ for OS, but the vorticity and streamwise directions rotate in the case of RS and RRS, whereas the gradient direction remains along $z$.}
\label{f:OS_RS_system_schematic}
\end{figure}

Figure \ref{f:OS_RS_system_schematic}a shows a schematic view of the dense suspension under consideration. As a reference, we will expose the suspension to the oscillatory shear (OS) protocol, which has been widely used in literature to investigate the time-dependent rheology of particulate suspensions. 
In our coordinate system, we consider particles subject to a time-periodic shear rate in the $y$ direction with $z$ being the gradient direction,
$\dot{\gamma}_{yz} = \gamma_0 \omega \cos(\omega t)$, 
as shown in Figure \ref{f:OS_RS_system_schematic}(b). 
To investigate a more complex flow, we propose to add to the $yz$ shear a {\it second} shear in the $xz$ plane with a non-zero phase difference to the first shear.

If the second shear is lagging to the first one by $\pi/2$ in phase,
$\dot{\gamma}_{xz} = \gamma_0 \omega \sin(\omega t)$,
the direction of the effective applied shear would rotate in the clockwise direction (see Figure \ref{f:OS_RS_system_schematic}c), which we call {\it rotary shear} (RS).
In practice, this might be realized, for example, in a cylindrical rheometer by oscillating the inner cylinder in both axial and circumferential directions.
The external rate-of-strain tensors corresponding to the OS and RS protocols are 

\begin{equation}
\setlength{\arraycolsep}{0pt}
\renewcommand{\arraystretch}{1.3}
    \mathbb{E}^\infty_\textrm{OS} = \left[
    \begin{array}{ccc}
        0 & \quad 0 & \ 0 \\
        0 & \quad 0 & \ \nicefrac{\dot{\gamma}_{yz}}{2} \\
        0 & \quad \nicefrac{\dot{\gamma}_{zy}}{2} & \  0 \\
    \end{array}  \right] , \quad
    \mathbb{E}^\infty_\textrm{RS} = \left[
    \begin{array}{ccc}
        0 & \quad 0 & \ \nicefrac{\dot{\gamma}_{xz}}{2} \\
        0 & \quad 0 & \ \nicefrac{\dot{\gamma}_{yz}}{2} \\
        \nicefrac{\dot{\gamma}_{zx}}{2} & \quad \nicefrac{\dot{\gamma}_{zy}}{2} & \  0 \\
    \end{array}  \right] .
\label{eq:strain_rate_tensor}
\end{equation}

In the RS protocol, the suspension is acted upon by shear of constant magnitude but varying direction. Importantly,  this leads to a flow that is \emph{not} time reversible. To nonetheless keep reversibility and also examine the effect of sudden shear reversal, we consider a rotary deformation where we reverse the flow direction \emph{and} the direction of rotation of the shear, which we call {\it reversible rotary shear} (RRS), see Figure \ref{f:OS_RS_system_schematic}(d). 
Specifically, we apply $\dot{\gamma}_{xz} \to -  \dot{\gamma}_{xz}$, $\dot{\gamma}_{yz} \to -  \dot{\gamma}_{yz}$, and $\omega \to - \omega$ after a rotation of angle  $\theta_0$ or after a time $t=\theta_0/\omega$. For consistency and ease of comparison, we keep the same time period ($T = 2\pi/\omega$) in OS, RS, and RRS such that the shear rate amplitude $\dot{\gamma} = \gamma_0\omega$ remains a small constant. Thus, the time-dependent shear rate in the RRS protocol is expressed as
\begin{equation}  
    \dot{\gamma}_{xz} =  \gamma_0 \hat\omega \sin(\hat\omega t), \quad
    \dot{\gamma}_{yz} =  \gamma_0 \hat\omega \cos(\hat\omega t),
\end{equation}
where $\hat\omega=\omega$ if $nT \leq t<(n+1/2)T$, and $\hat\omega=-\omega$ if $(n+1/2)T \leq t< (n+1)T$, for $n \in \mathbb{N}$. 

The motion of a rigid particle in a particle suspension is governed by the Newton-Euler equations,
\begin{equation}\label{eq:ne}
\begin{array}{ll}
    {\bf F}_i  = m_i \frac{d {\bf u}_i}{dt},\\[8pt]
    {\bf T}_i = {\bf I}_i \frac{d{\bf w}_i}{dt} + {\bf w}_i \times ({\bf I}_i {\bf w}_i),
\end{array}
\end{equation}
where ${\bf F}_i$ and ${\bf T}_i$ are, respectively, the total force and torque exerted on particle $i$ of mass $m_i$ and moment of inertia tensor ${\bf I}_i$, and ${\bf u}_i$ and ${\bf w}_i$ its translational and rotational velocities, respectively.
For spheres, rotation has no effect on the translational dynamics of the suspension, thus we omit the associated degrees of freedom from now on.
When the suspension is dense, long-range hydrodynamic interactions may also be negelcted \citep{seto2013discontinuous, mari2014shear}, and the total force on a particle is simply a sum of hydrodynamic forces, such as the Stokes drag $({\bf F}_i^S)$ and pairwise, short-range lubrication $({\bf F}_i^L)$, and any non-hydrodynamic forces, including contact forces $({\bf F}_i^C)$ and friction. Specifically, we write,\begin{equation}\label{eq:force}
    {\bf F}_i = {\bf F}_i^S + \sum_j^{N_l} {\bf F}^L_{ij} + \sum_j^{N_c} {\bf F}^C_{ij},
\end{equation}
where $N_l$ and $N_c$ are the numbers of lubricating and contact pairs. The expression for the individual terms on the right-hand side of the above equation are taken from the hybrid lubrication/granular dynamics (HLGD) model derived for dense particle suspensions in low Reynolds number flow \citep{cheal2018rheology, ge2020implementation}. Their detailed expressions are given in Appendix \ref{app_sec:forces}.

Finally, we note that there are several model parameters in our system, including the particle radius $a$, density $\rho$, stiffness $k_n$, dynamic viscosity of the fluid $\eta_0$, and shear rate $\dot\gamma$. For the model output
to correspond physically to a dense suspension of inertialess, rigid particle, we require the Stokes number St $=\rho \dot{\gamma} a^2/\eta_0 \ll 1$ and a stiffness-scaled shear rate $\hat{\dot{\gamma}} = \dot{\gamma}a/\sqrt{k_n/(\rho a_1)} \ll 1$. Throughout this work, we have St $ \approx 10^{-2}$ and $\hat{\dot{\gamma}} \approx 10^{-4}$. See \citet{ge2020implementation} for more details.

\subsection{Quantities of interest}\label{subsec:qoi}

To examine the rheology of the suspension, we consider the bulk stress tensor, defined as
\begin{equation}
    \sigma = 2\eta_0\mathbb{E}^\infty + \frac{1}{V} \left( \sum_i^N \mathbb{S}_i^S + \sum_{j>i}^{N_l} \mathbb{S}_{ij}^L + \sum_{j>i}^{N_c} \mathbb{S}_{ij}^C \right),
\end{equation}
where $V = L_xL_yL_z$ is the volume of the simulation box (c.f.~Figure \ref{f:OS_RS_system_schematic}a), and the $\mathbb{S}$'s denote the stresslets due to the various interactions. 
Specifically, the single-body hydrodynamic stresslet ($\mathbb{S}^S$), the pairwise lubrication stresslet ($\mathbb{S}^L$), and the contact-force stresslet ($\mathbb{S}^C$) are given as
\begin{subequations}
    \begin{equation}
        \mathbb{S}_i^S = \left(20\pi\eta_0 a_i^3/3 \right)\mathbb{E}^\infty,  
    \end{equation}
    \begin{equation}
        \mathbb{S}_{ij}^L = {\bf F}_{ij}^L({\bf r}_j- {\bf r}_i),    
    \end{equation}
    \begin{equation}
        \mathbb{S}_{ij}^C = {\bf F}_{ij}^C({\bf r}_j- {\bf r}_i),    
    \end{equation}
\end{subequations}
where, $i$ and $j$ are particle indices. In steady shear flow, the viscosity is simply $\eta = \sigma_{yz}/\dot{\gamma}$, where $\sigma_{yz}$ is the shear stress component of the output stress tensor, and $\dot{\gamma}$ is the shear rate. 
In the unsteady shear case, the concept of complex viscosity is adopted, which includes both viscous and elastic contributions. It is customary to express the complex viscosity $\eta^*$ as

\begin{equation} 
\eta^* = \eta' - i\eta'',
\end{equation}
where $\eta'$ denotes the dynamic viscosity, and $\eta''$ the non-dissipating elastic contribution \citep{mewis2012colloidal}.
The complex viscosity, $\eta^*$ can be calculated using Fourier analysis for the OS protocol. The shear stress along the flow direction within a cycle can be decomposed in a Fourier series,
\begin{equation}
    \sigma_{yz}(t) = \gamma_0 \sum_{n=1}^{\infty}(G'_n \sin(n\omega t) + \omega\eta'_n \cos(n\omega t)), 
\end{equation}
where $G'_n$ and $\eta'_n$ denote the $n$th elastic and viscous coefficients, respectivly. 

\begin{figure}
  \centering
  \begin{subfigure}{0.45\textwidth}
      \centering
      \includegraphics[width=\textwidth]{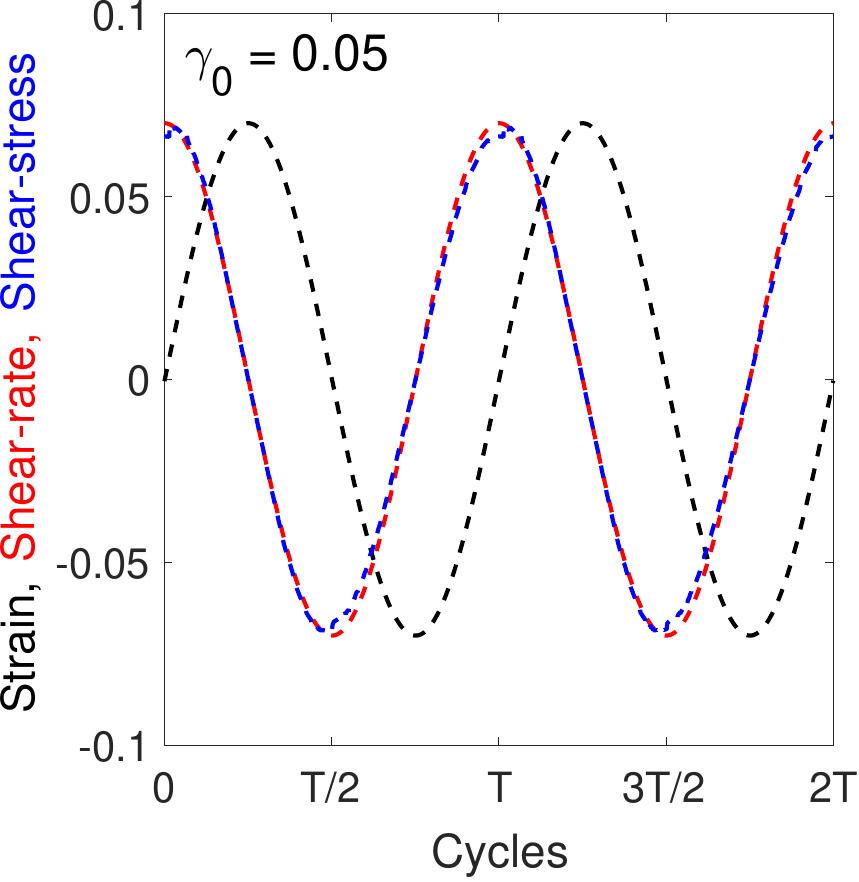} 
      \caption{}
  \end{subfigure}
  \hspace{0.218118in}
  \begin{subfigure}{0.45\textwidth}
      \centering
      \includegraphics[width=\textwidth]{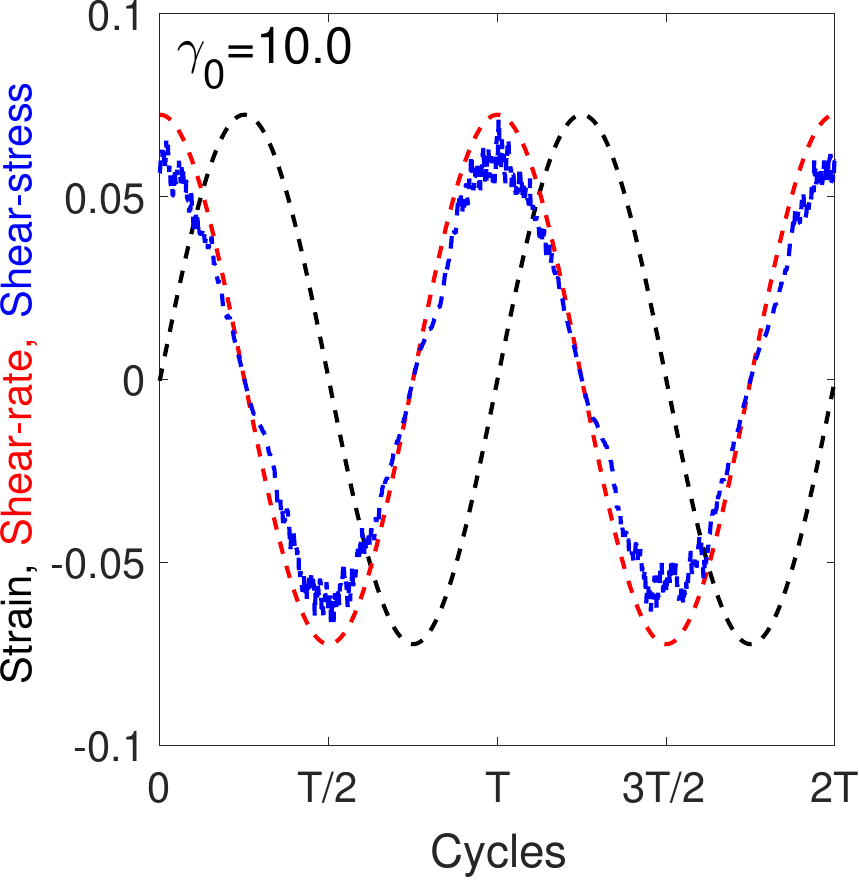} 
      \caption{}
  \end{subfigure}
\caption{Stress-strain evolution for OS (a) $\gamma_0=0.05$, (b) $\gamma_0=10.0$. The signals are normalized for comparison. Here, $T$ represents the period of a shear cycle.} 
\label{f:stress_strain}
\end{figure}

In the case of RS and RRS, the flow-vorticity plane rotates with respect to the gradient direction. Therefore we need to use tensor rotation to calculate the stress tensor in the frame of reference rotating with the imposed shear,
\begin{equation}
    \sigma' = R \sigma R^T,
\end{equation}
where $R$ is the rotation matrix (c.f. Eq.~\ref{eq:r_matrix}), and $\sigma$ and $\sigma'$ are the stress tensor in the laboratory and rotating frame respectively.   
The normal and shear stress components are given as 
\begin{equation}\label{eq:stress_rot}
    \begin{array}{ll}
        \sigma'_\text{ff} &= \sigma_\text{xx}\sin^2(\theta) + ( \sigma_\text{xy} + \sigma_\text{yx})\sin(\theta)\cos(\theta) + \sigma_\text{yy}\cos^2(\theta),\\
        \sigma'_\text{gg} &= \sigma_\text{zz},\\
        \sigma'_\text{vv} &= \sigma_\text{xx}\cos^2(\theta) - (\sigma_\text{yx}+\sigma_\text{xy})\sin(\theta)\cos(\theta) + \sigma_\text{yy}\sin^2(\theta),\\
        \sigma'_\text{fg} &= \sigma_\text{xz}\sin(\theta)+\sigma_\text{yz}\cos(\theta),\\
        \sigma'_\text{vg} &= \sigma_\text{xz}\cos(\theta)-\sigma_\text{yz}\sin(\theta),\\
    \end{array}
\end{equation}
where $\theta = \omega t$ is the angle of rotation of the flow-vorticity plane, $\sigma'_\text{ff}$, $\sigma'_\text{gg}$, and $\sigma'_\text{vv}$ are, respectively, the normal stresses along the instantaneous flow, gradient and vorticity directions, and $\sigma'_\text{fg}$, and $\sigma'_\text{vg}$ are the viscous and elastic components of the shear stress in the rotating frame. 
Note that the normal stresses in the lab frame do not contribute to dissipation, while the shear stress in the lab frame appears in the normal stresses in the rotating frame. 
The suspension viscosity under RS and RRS can be calculated as,
\begin{equation}
    \eta' = \sigma'_\text{fg}/\gamma_0\omega,
\end{equation}
\begin{equation}
    \eta'' = -\sigma'_\text{vg}/\gamma_0\omega,
\end{equation}
\begin{equation}
    \eta^* = \sqrt{\eta'^{2}+\eta''^{2}}.
\end{equation}
In this work, we report the relative complex viscosities, $\eta_r^* = \eta^*/ \eta_0$. In our particulate system, the stress response is closely in phase with the shear rate irrrespective of the strain amplitude, as shown in Figure \ref{f:stress_strain}. Therefore, the complex viscosity has a dominant viscous contribution, and complex viscosity and dynamic viscosity are typically not distinguishable in the plots. 

In order to understand the non-Newtonian nature of the flow, we also examine the first, $N_1$, and second, $N_2$, normal stress differences, defined as
\begin{equation}\label{eq:stress_diff}
    \begin{array}{ccc}
        N_1 &=& \sigma'_\text{ff} - \sigma'_\text{gg},\\
        N_2 &=& \sigma'_\text{gg} - \sigma'_\text{vv}.
    \end{array}
\end{equation}
For the OS protocol, the stress terms in the above equation are simply $\sigma'_\text{ff} = \sigma_\text{yy}$, $\sigma'_\text{gg} = \sigma_\text{zz}$, $\sigma'_\text{vv} = \sigma_{xx}$. For the RS and RRS protocol, on the other hand, 
the normal stresses are obtained from Eq. \ref{eq:stress_rot}.

To investigate the suspension dynamics we calculate the mean squared displacement (MSD) of the particle ensemble,
\begin{equation}\label{eq:msd}
\big\langle [\Delta r (\gamma_t)/d]^2 \big\rangle = 6 D_\text{eff} \gamma_t, 
\end{equation}
where $\Delta r$ is the particle displacement, $d=2a$ the particle diameter, $\langle \cdot \rangle$ an average over all particles and the time, $D_\text{eff}$ the effective diffusivity,  and $\gamma_t$ the total strain: $\gamma_t = |\dot{\gamma}|t$, for SS; $\gamma_t = 4 \gamma_0 n$, for OS; $\gamma_t = 2 \pi \gamma_0 n$, for RS, and RRS, for $n$ number of shear cycles. All the simulations are performed for at least $\gamma_t = 300$ strain units. 

To understand the evolution of the suspension microstructure we calculate the coordination number $(Z)$, i.e.\ the average number of particles in contact with any particle in the suspension, 
as well as the pairwise distribution of particles $g(h,\theta)$ which provides information on the relative distribution of particle pairs in the suspension in terms of their radial separation $(h)$ and relative orientation $(\theta)$ when projected on the shear plane. Here, $h\leq0$ shows particle pairs in contact, and $h>0$ shows that particles are separated. The angle $\theta$ is calculated with respect to the shear direction, where $\theta \in [0,\pi/2]$ and $[\pi, 3\pi/2]$ represent the two extensional quadrants, and $\theta \in [\pi/2, \pi]$ and $[3\pi/2, 2\pi]$ the two compressional quadrants.

\subsection{Problem set-up}\label{subsec:psetup} 
In our simulations, we consider a bidisperse suspension of $500$ particles with a size ratio of $1.4$ and particle number such that the suspension has an equal volume fraction of large and small particles. The computational domain is a cubic box with Lees-Edwards boundary conditions \citep{lees1972computer}. The initial state is obtained for each simulation by preshearing a random configuration with a constant shear rate for a total strain of $\dot{\gamma}t = 40$. Note that this results in an initially anisotropic microstructure with many particles in contact. The suspension is then subject to the different shear protocols under consideration (OS, RS, and RRS) and data is collected over at least 200 accumulated strains. 
The control parameters are the volume fraction of the solid particles $\phi$, the friction coefficient for particle-particle contact $\mu$, and the strain amplitude $\gamma_0$. 
In this work, we present results for $\phi =$ 0.40, 0.50, and 0.55, and $\mu_c=$ 0.0, 0.2, and 0.5 to focus on the dense regime, where frictional contact is most prominent. The strain
$\gamma_0$ is varied from 0.05 to 10, while keeping the maximal shear rate $\gamma_0\omega$ constant, to explore both small and large-amplitude oscillatory/rotary shear.
For RRS, the shear direction is reversed every half a cycle ~i.e. after a rotation of $\pi$ radians, if not specified otherwise.

\begin{figure*}
  \centering
    \includegraphics[width=0.8\textwidth]{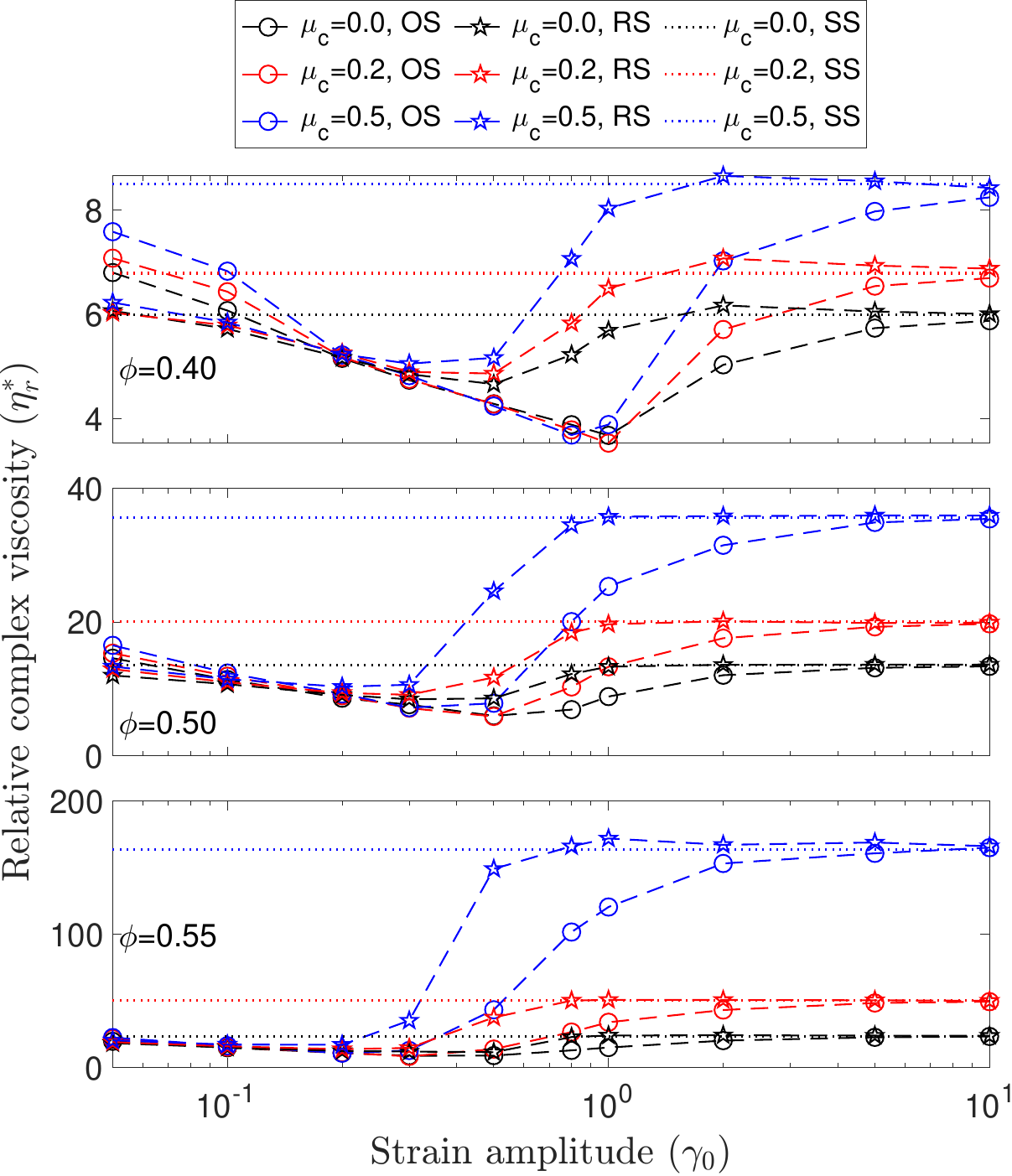} 
\caption{(a) Complex viscosity for OS (circles), RS (stars), SS (dotted lines). Black lines $\mu_c=0.0$, red lines $\mu_c=0.2$, blue lines $\mu_c=0.5$.}
\label{f:comp_viscosities_OS_RS}
\end{figure*}

\section{Results}\label{sec:results}
\subsection{Rheology}\label{subsec:rheology}

First, we report the main rheological observables: the complex viscosity and normal stress differences. 
Figure~\ref{f:comp_viscosities_OS_RS} shows the suspension relative complex viscosity, $\eta_r^*$, as a function of the strain amplitude $\gamma_0$ at different volume fractions $\phi$ and friction coefficients $\mu_c$, subject to either OS (circle), RS (star) or SS (dotted). The main qualitative observation is a non-monotonic dependence of $\eta_r^*$ on $\gamma_0$: specifically, $\eta_r^*$ first decreases with the applied strain until it reaches a minimum at an intermediate strain $\gamma_{0,m}$ (denoted hereafter the minimum strain amplitude), then increases and converges to the steady-shear viscosity at large $\gamma_0$ (reported by dotted lines in the figure).
In OS, a non-monotonic complex viscosity with respect to the strain amplitude has long been observed \citep{breedveld2001shear, bricker2006oscillatory, bricker2007correlation}, and can be explained by the shear-induced microstructure and its evolution under repeated flow reversal \citep{ge2022rheology}.
The observation of a similar viscous response in RS suggests that rotating the shear while keeping a constant shear rate has, on average, a similar effect on the suspension microstructure as oscillating the shear rate in one direction, since the suspension rheology ultimately depends on its microstructure. Despite the qualitative similarity of the viscous responses, the specific value of the minimum viscosity is higher in RS as compared to OS. As we will discuss later, this can be attributed to the absence of microstructure breaking in RS, resulting in the continuous resistance to the flow unlike that occurs within a shear cycle in OS.

The specific value of the minimum strain amplitude $\gamma_{0,m}$ depends on the volume fraction $\phi$, the shearing protocol, and the friction coefficient $\mu_c$ in a complex way. Generally, $\gamma_{0,m}$ is smaller under the RS protocol than under the OS protocol. At $\phi=0.40$, the minimum viscosity is attained at $\gamma_{0,m} \approx 1$ for all three values of the friction coefficient $\mu_c$ in OS \citep[as also observed in the experiments by][]{bricker2006oscillatory,bricker2007correlation}, 
whereas the minimum moves to smaller $\gamma_0$ values in RS. Varying $\mu_c$ has only a weak effect on $\gamma_{0,m}$, slightly shifting $\gamma_{0,m}$ to smaller values as it increases.    
With the increase in volume fraction $\phi$, $\gamma_{0,m}$ decreases and the dependence of $\gamma_{0,m}$ on the shearing protocol reduces. Since $\gamma_{0,m}$ is a purely geometric parameter in the absence of attractive interparticle interactions \citep{corte2008random, ge2021irreversibility}, 
its dependence on $\mu_c$ and the shearing protocol indicates a subtle interplay of the surface roughness (modeled here by $\mu_c$) and the microstructure formed under shear \citep{lemaire2023rheology}.

\begin{figure*}
  \centering
  \includegraphics[scale=0.45]{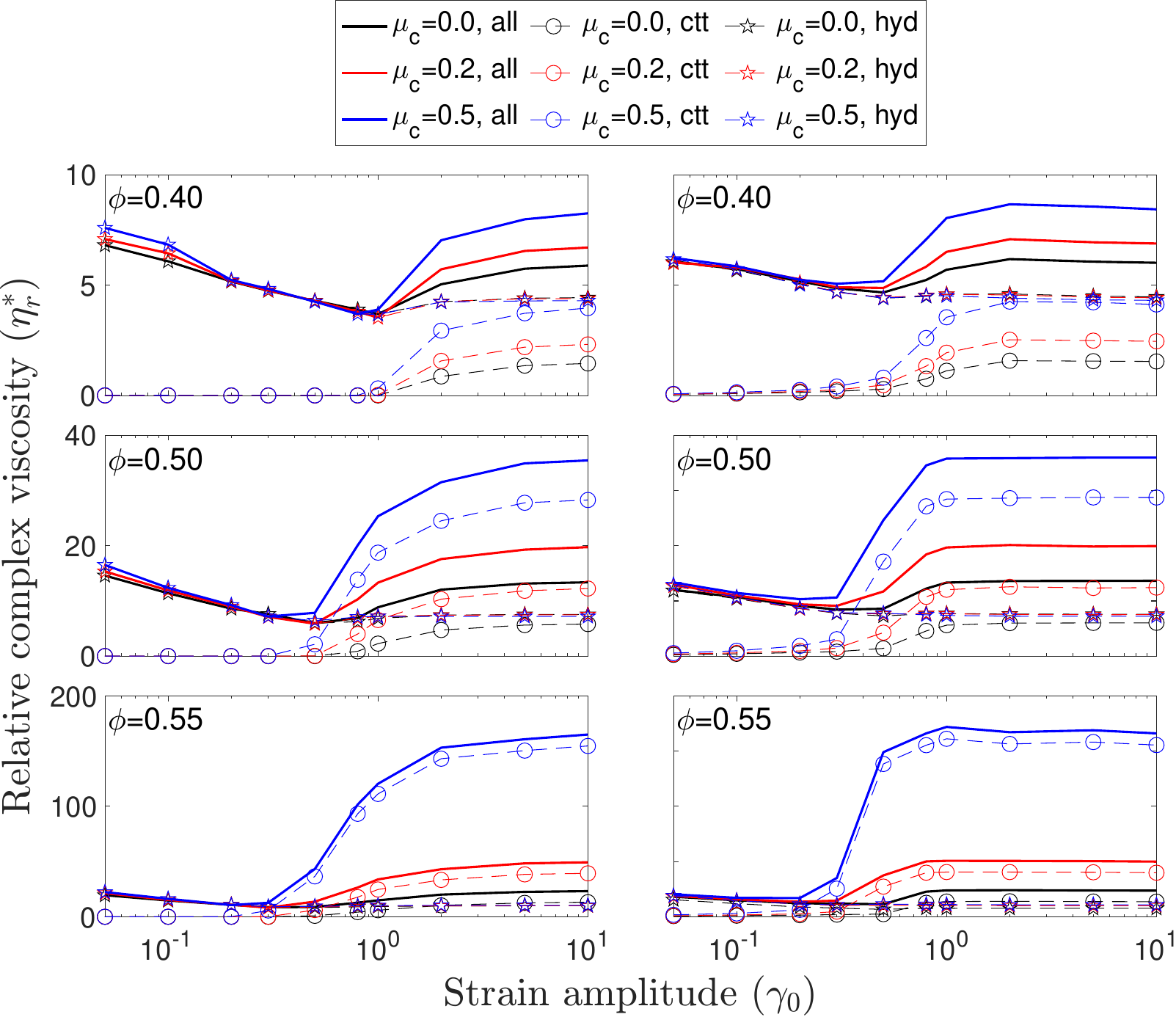} 
  \caption{Viscosity budget, total stress (solid lines), contribution from contact (circles), contribution from hydrodynamic (stars) for OS (left column) and RS (right column). Friction coefficient, $\mu_c=0.0$ (black lines), $\mu_c=0.2$ (red lines), $\mu_c=0.5$ (blue lines).  $\phi=0.40$ (top row), $\phi=0.50$ (middle row), $\phi=0.55$ (bottom row).}
  \label{f:stress_budget_OS_RS}
\end{figure*}
\begin{figure}
    \centering
    \begin{subfigure}{0.45\textwidth}
        \centering
        \includegraphics[width=\textwidth]{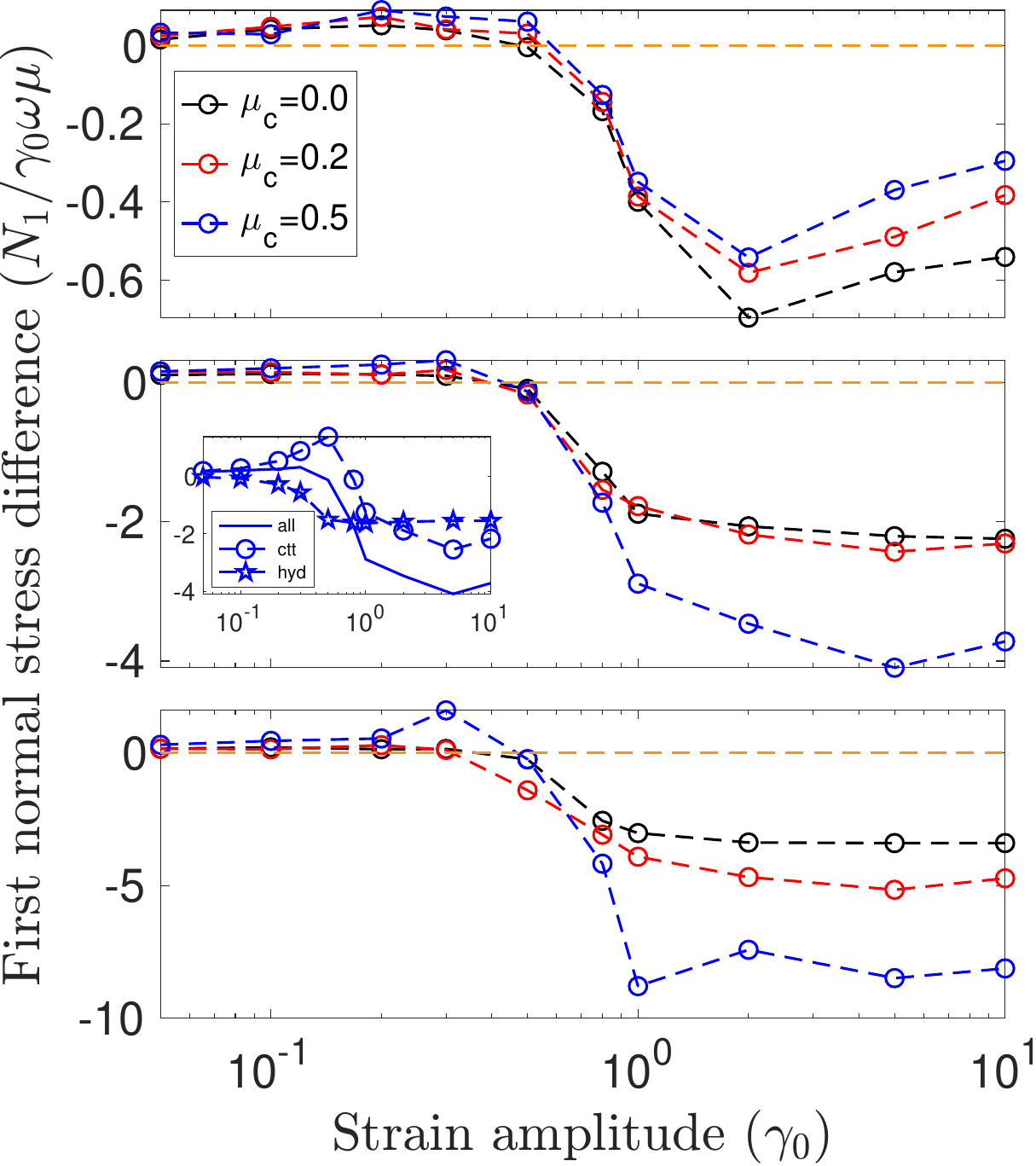}
         \caption{}
         \label{f:N2_os}
    \end{subfigure}
    \begin{subfigure}{0.45\textwidth}
        \centering
        \includegraphics[width=\textwidth]{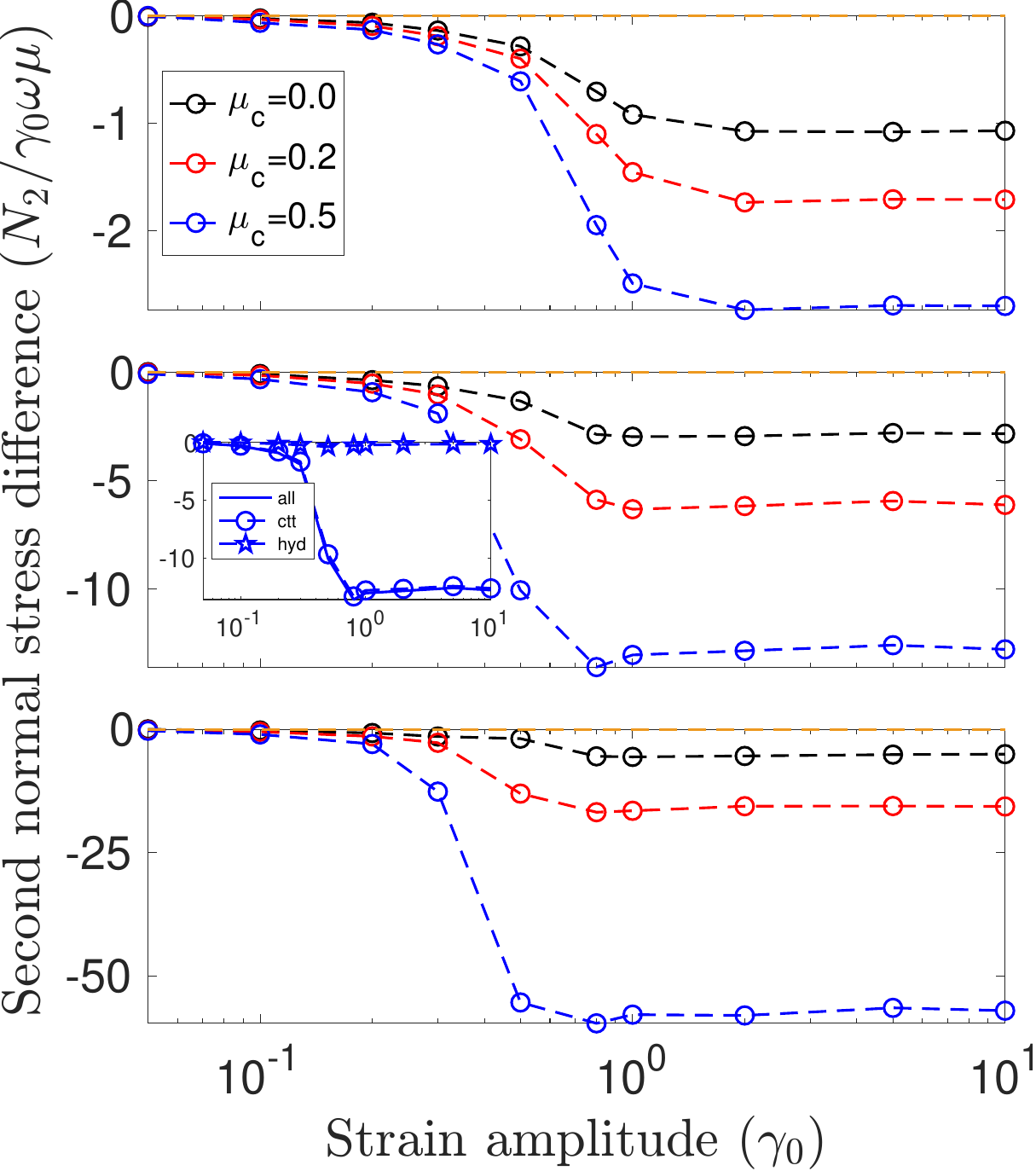} 
         \caption{}
         \label{f:N2_rs}
    \end{subfigure}
    \caption{The first (a) and second (b) normal stress differences under RS at three volume fractions: (top row) $\phi=0.40$, (middle row) $\phi=0.50$, and (bottom row) $\phi=0.55$. The inset figures show the individual contribution from the contact and hydrodynamic forces at $\phi=0.50$ and $\mu_c=0.5$. The yellow dashed line shows the zero line for reference.}
    \label{f:stress_diff_rs}
\end{figure}

\begin{figure}
    \centering
    \begin{subfigure}{0.45\textwidth}
        \centering
        \includegraphics[width=\textwidth]{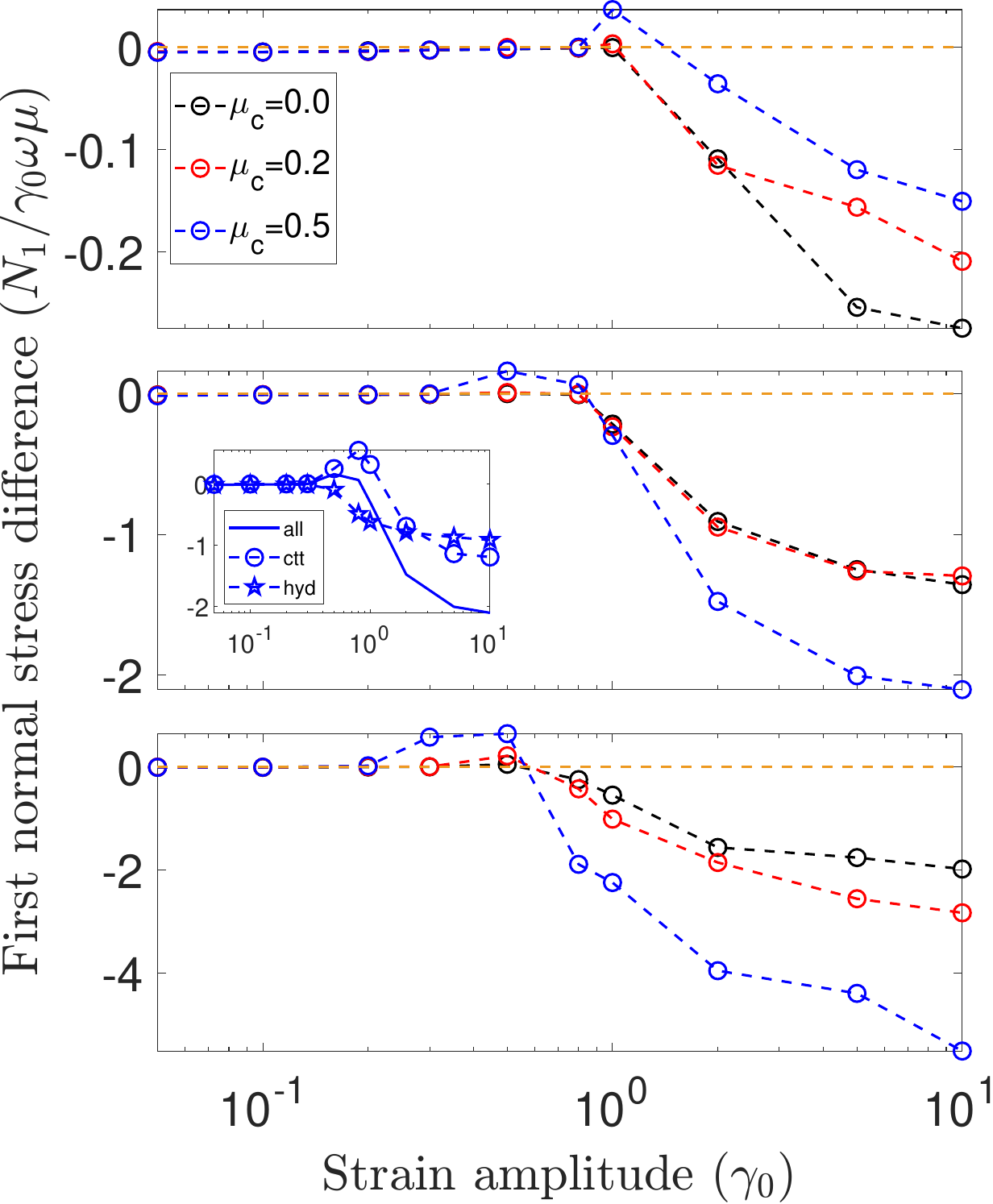}
         \caption{}
         \label{f:N1_os}
    \end{subfigure}
    \begin{subfigure}{0.45\textwidth}
        \centering
        \includegraphics[width=\textwidth]{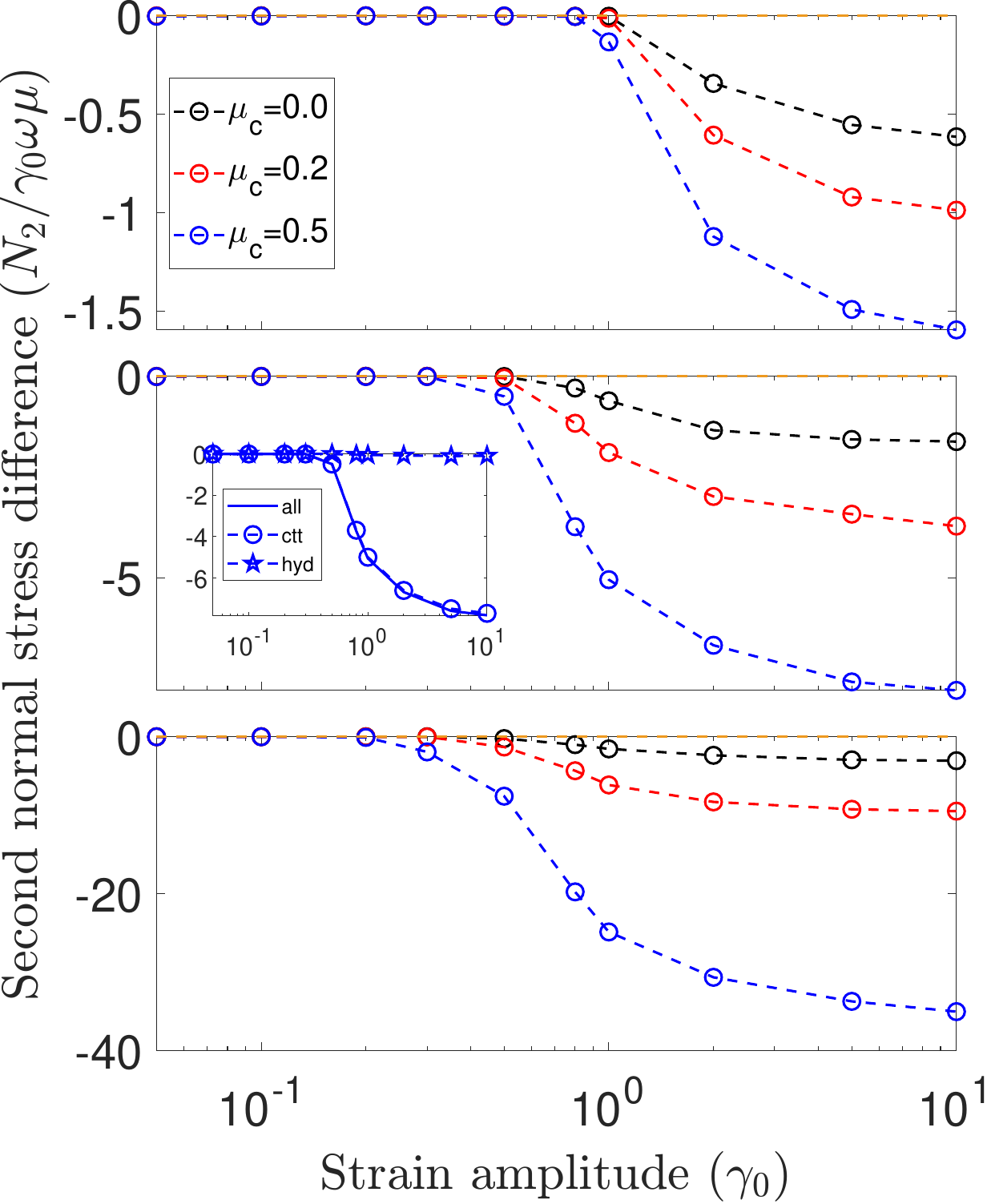} 
         \caption{}
         \label{f:N1_rs}
    \end{subfigure}
    \caption{The first (a) and second (b) normal stress differences under OS at three volume fractions: (top row) $\phi=0.40$, (middle row) $\phi=0.50$, and (bottom row) $\phi=0.55$. The inset figures show the individual contribution from the contact and hydrodynamic forces at $\phi=0.50$ and $\mu_c=0.5$. The yellow dashed line shows the zero line for reference.
    }
    \label{f:stress_diff_os}
\end{figure}

To better understand the non-monotonic variation of the viscosity, we show the contributions from the hydrodynamic and non-hydrodynamic (contact) stresses to $\eta_r^*$, obtained from the total stress, for all values of $\phi$ and $\mu_c$ under OS or RS in Figure~\ref{f:stress_budget_OS_RS}. The decomposition reveals two rheological regimes. First, at low $\gamma_0$, the stress is always dominated by the hydrodynamic contribution (mainly lubrication), which is independent of $\mu_c$ and decreases with increasing $\gamma_0$, leading to the reduction of the total viscosity at moderate $\gamma_0$.
Second, from moderate to high $\gamma_0$, the contact stress also contributes to the rheology (especially at large $\phi$ or $\mu_c$), and its role increases with $\gamma_0$.
Note that, new to the RS protocol, the contact contribution is already nonzero for $\gamma_0 < \gamma_{0,m}$.
This indicates the presence of particle contact at an earlier strain amplitude and explains why the minimal $\eta_r^*$ is higher in magnitude and occurs at smaller $\gamma_{0}$ in RS than in OS. For the low strain amplitudes, $\mu_c$ has an almost negligible effect in either protocol as friction does not affect the hydrodynamic forces and, at smaller amplitudes, particles should not be in contact. On the other hand, $\mu_c$ has a more prominent effect on $\eta_r^*$ at large strain amplitudes, when the viscosity increases dramatically with $\mu_c$ for the same $\phi$. This is expected as the presence of friction tends to enhance the importance of contacts. On the other hand both the contact and hydrodynamic stresses increase with $\phi$, increasing the overall complex viscosity.

In addition to the complex viscosity, the first ($N_1$) and second ($N_2$) normal stress differences (NSD) are typically used to characterize the rheology of suspensions. The cycle-averaged values of $N_1$ and $N_2$ are displayed in Figure \ref{f:stress_diff_rs} and \ref{f:stress_diff_os} as a function of $\gamma_0$ for different values of $\phi$ and $\mu_c$ for RS and OS, respectively. While both $N_1$ and $N_2$ are negligible at small $\gamma_0$, they become finite and negative when $\gamma_0 > \gamma_{0,m}$. The negligible $N_1$ at smaller $\gamma_0$ suggests that the pair distribution function characterizing the suspension microstructure is nearly fore-aft symmetric \citep{morris2009review}; its finite values at larger $\gamma_0$ indicates the breaking of such symmetry. 
Another observation is that $N_1$ is slightly positive (more so at larger $\mu_c$) at $\gamma_0 \approx \gamma_{0,m}$. As discussed in \cite{ge2022rheology}, this is due to the more frequent particle collisions (resulting in compressive normal stresses) in the velocity gradient direction than in the flow direction; c.f.~Figure \ref{f:stress_diff_os} inset and Figure 7 in \cite{lemaire2023rheology}. Here, our results further suggest that the effect can be enhanced by surface roughness (i.e.~larger $\mu_c$).
As for $N_2$, it is approximately zero at small strain amplitudes, but increases (in magnitude) sharply around $\gamma_{0,m}$, until it saturates to its corresponding steady shear value. A negative value of $N_2$ is widely observed in the literature \citep{guazzelli2018rheology}, which is attributed to the dominant contribution from contact stresses that occur mostly in the shear plane, i.e., there are more collisions in the velocity gradient direction than in the vorticity direction. The normal stress differences for RS are consistent with the behavior observed in our OS protocol, shown in Figure \ref{f:stress_diff_os}, and also in the previous studies on OS.



Moreover, we note that both $|{N_1}|$ and $|{N_2}|$ increase with the friction coefficient $\mu_c$ at higher applied strain amplitude $\gamma_0$ for both OS and RS (with an exception at 40\%).
Because the NSDs are mainly determined by contact forces at higher $\gamma_0$, the presence of friction enhances the effect of contacts. However, the strain amplitude at which $N_2$ becomes finite is only weakly sensitive to the value of $\mu_c$ and $\phi$, as the (absolute) slopes of the curves barely increase with $\mu_c$ or $\phi$ near the onset of $N_2$ (c.f.~the relatively larger difference in OS, Figure \ref{f:stress_diff_os}b).
Finally, we note that recent work by \cite{ge2022rheology} has shown that certain rheological features, such as a minimum complex viscosity and the onset of second normal stress difference, occur at the critical strain amplitude at which the suspension undergoes a reversible-irreversible transition (RIT). Since we observe similar rheological features in RS, we examine the dynamical response of our suspensions in the next section.

\subsection{Dynamics}

\begin{figure*}
  \centering
  \begin{subfigure}{0.48\textwidth}
      \centering
      \includegraphics[width=\textwidth]{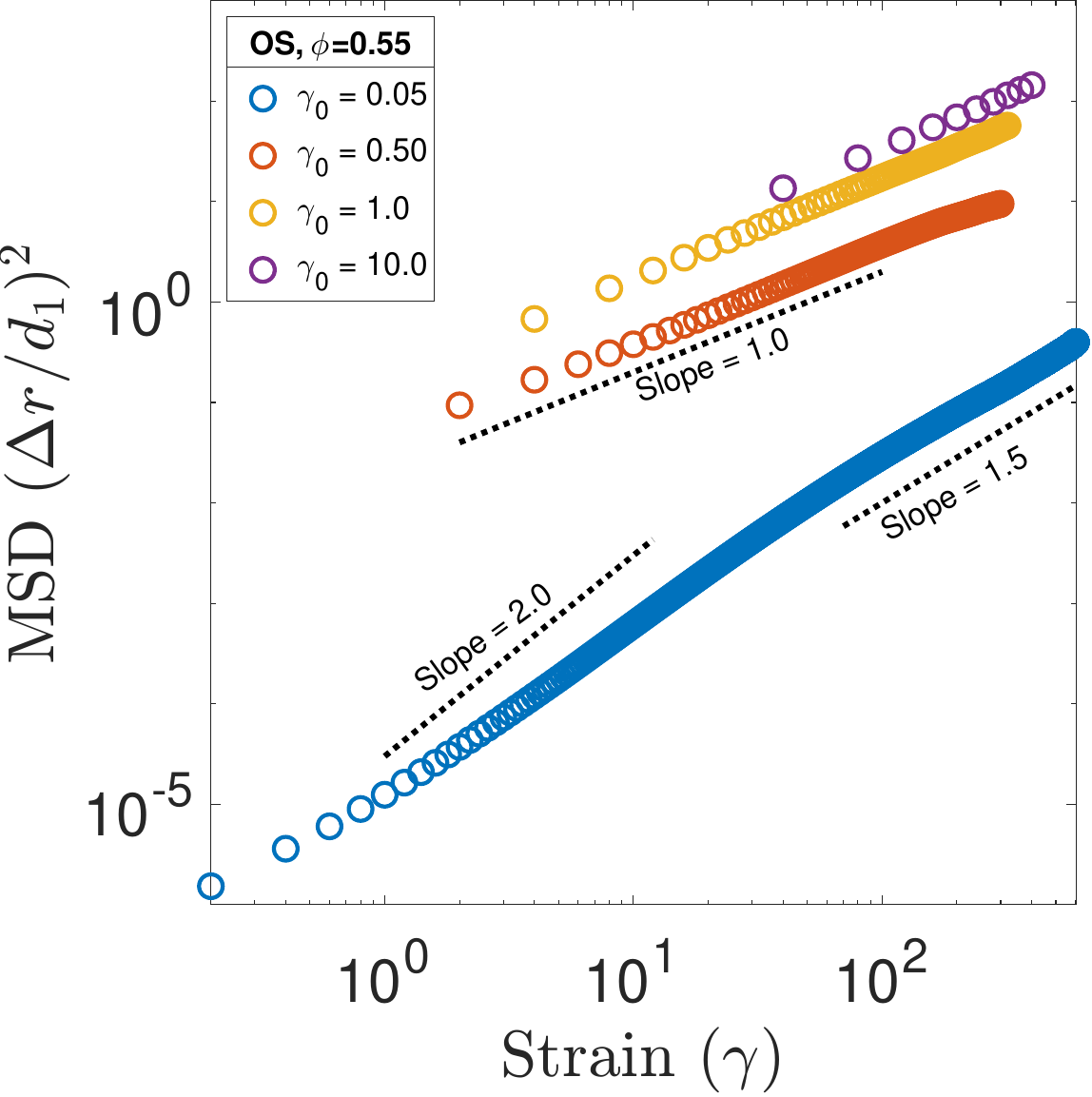} 
      \caption{}
  \end{subfigure}
  \hspace{0.3cm}
  \begin{subfigure}{0.48\textwidth}
      \centering
      \includegraphics[width=\textwidth]{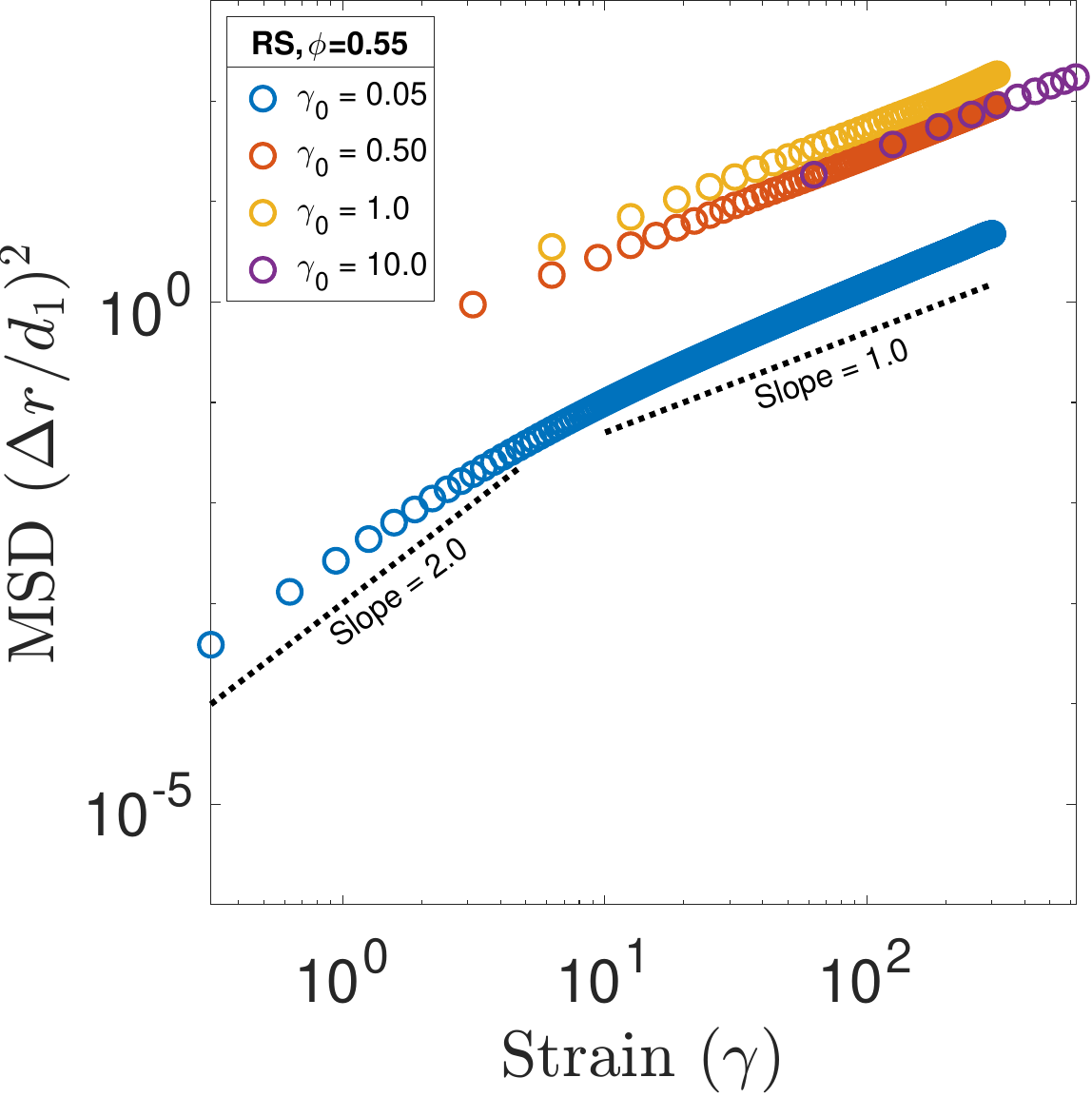} 
      \caption{}
  \end{subfigure}
  \caption{Mean square displacements (MSD) for OS (a) and RS (b) at different $\gamma_0$ for $\phi=0.55$ and $\mu_c= 0.5$. The black lines are for reference to show the slopes.}
  \label{f:MSD}
\end{figure*}

In this section, we investigate the particle dynamics in RS, comparing it with the standard OS protocol. 
In the case of OS, previous works have shown that periodically sheared suspensions can display a dynamical phase transition called reversible-irreversible transition (RIT), or absorbing-phase transition \citep{pine2005chaos, corte2008random, menon2009universality}. 
Specially, there is a $\phi$-dependent critical strain amplitude $\gamma_{0,c}$ below which the suspension can fall into a reversible ``absorbing" state, i.e., all particles returning to their original positions after each strain cycle. On the other hand, for strain amplitudes larger than $\gamma_{0,c}$, the suspension may remain in an irreversible, fluctuating state, where particles do not return to their original positions and display biased random walk-like dynamics if observed at the end of each strain cycle \citep{pine2005chaos}. 

To investigate the presence of RIT, we examine the mean squared displacements (MSD) of the suspended particles, as shown in Figure \ref{f:MSD}. 
For OS, our results show that at large strain amplitudes the MSD scales linearly with the total strain at large strains, which indicates diffusive dynamics. 
On the other hand, at lower strain amplitudes, the MSD increases faster than linearly; however, its value is much smaller and the dynamics are practically negligible. As for RS, the MSD scales linearly for all simulated strain amplitudes, see e.g.~data for $\gamma_0=0.05$ in Figure \ref{f:MSD}.

\begin{figure*}
  \centering
  \includegraphics[width=0.8\textwidth]{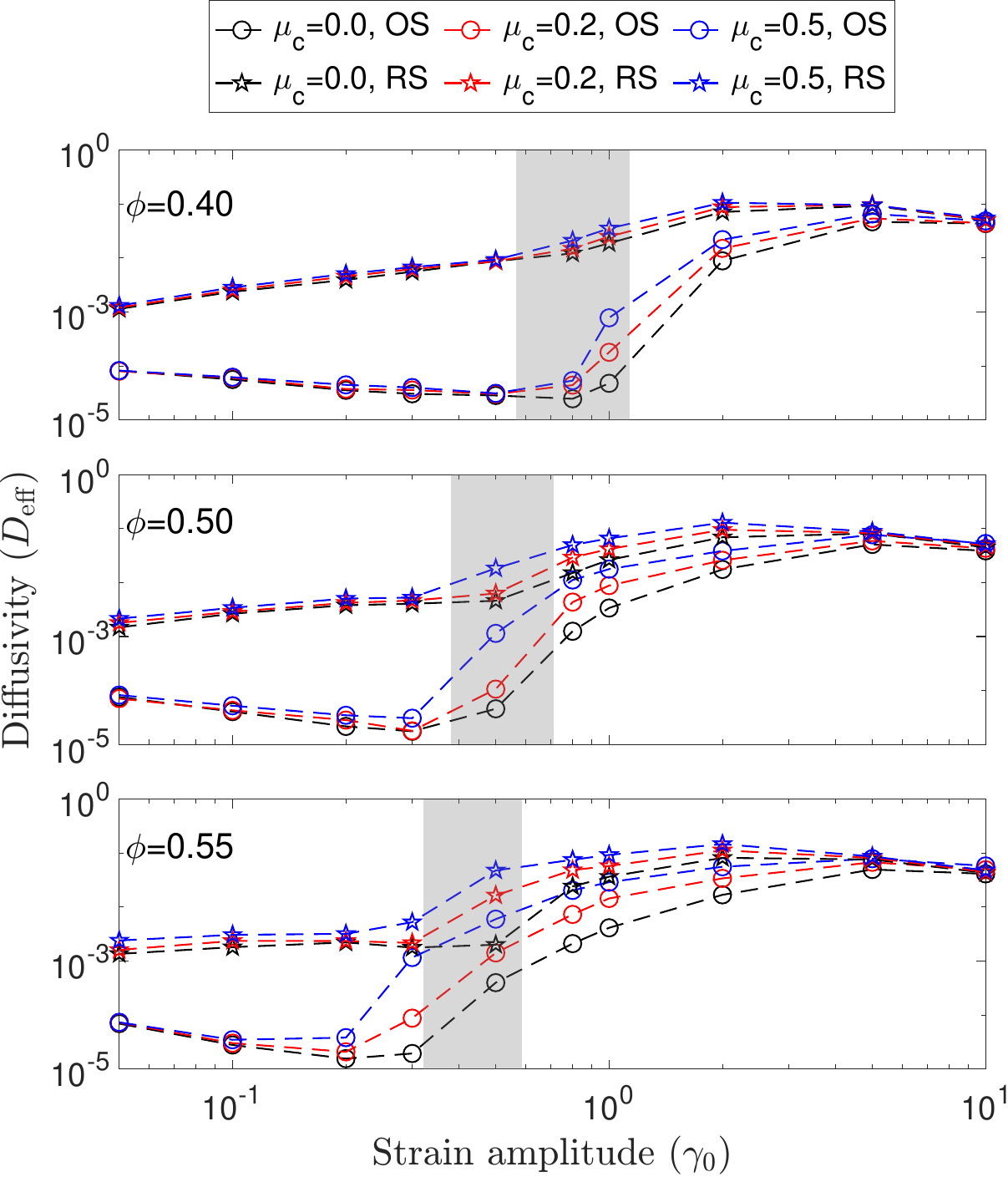} 
  \caption{Effective diffusivity ($D_\textrm{eff}$) vs $\gamma_0$ for OS and RS. The shaded region shows the region of prediction of the critical amplitude from the empirical scaling, $\gamma_{0,c} = C\phi^{-\alpha}$, where $C = 0.14 \pm 0.03$, and $\alpha = 1.93 \pm 0.14$ \citep{pine2005chaos}.
  }
  \label{f:deff_amp}
\end{figure*}

We calculate the effective diffusivity $D_\textrm{eff}$ for both OS and RS protocols from the linear fit of the MSD data similar to \cite{pine2005chaos}; see Figure \ref{f:deff_amp}. 
Here, $D_\textrm{eff}$ is calculated after the suspension has been sheared for $300$ strain units. 
For OS, the dynamics clearly display two regimes and a transition in between, consistent with the RIT; however, we cannot rigorously determine the critical strain amplitude $\gamma_{0,c}$ at which the transition occurs. As an alternative, we plot the experimentally fitted $\gamma_{0,c}$ by \citet{pine2005chaos}, $\gamma_{0,c} = C\phi^{-\alpha}$, where $C = 0.14 \pm 0.03$, and $\alpha = 1.93 \pm 0.14$ (shaded regions in Figure \ref{f:deff_amp}). The fitted values (based on experimental data at $\phi \leqslant 0.4$) appears to be slightly higher than our simulations, particularly at $\phi > 0.4$ and $\mu_c >0$. This could be due to differences between the experimental system and our numerical model or uncertainties in the fitting procedures. Nevertheless, the general trend of an earlier transition at higher $\phi$ is consistently observed.

For the suspension undergoing RS, $D_\textrm{eff}$ increases gradually with $\gamma_0$ and is not negligible even at the smallest strain amplitudes, in stark contrast to the dynamics in OS. 
As an example, note that $D_\textrm{eff}$ can be about 2 orders of magnitude larger than for the corresponding OS case at the same strain below $\gamma_{0,m}$. 
Although the rate of increase of $D_\textrm{eff}$ with respect to $\gamma_{0}$ seems to be higher at around $\gamma_{0,m}$ (c.f.~the $\phi=0.55$ case in the figure), the dynamics are always diffusive even at the smallest $\gamma_0$ (c.f.~Figure \ref{f:MSD}b). 
Therefore, we show that a non-monotonic rheological response under periodic shear, with a minimal viscosity at an intermediate $\gamma_0$, does not necessarily imply the existence of RIT.
Another interesting observation is that, although $D_\textrm{eff}$ depends on $\mu_c$ in the diffusive regime in OS (and to a less extent over the same range of $\gamma_0$ in RS), the dependence diminishes at larger $\gamma_0$ in the irreversible regime. 
As irreversible dynamics arise from non-hydrodynamic particle contacts, we examine their effect on the microstructure next.

\subsection{Microstructure}\label{subsec:microstructure}

Non-Newtonian stresses are generally caused by an anisotropic microstructure. Because hard spheres do not  deform are isotropic in shape, the microstructure is entirely dictated by the spatial arrangement in the suspension. To find the origin of the differences in the suspension dynamics between the sudden shear reversal (OS) and the gradual shear reversal (RS) protocols, we first examine the particle coordination number ($Z$), defined as the average number of particles in contact with one particle in suspension.

\begin{figure*}
\centering
  \centering
  \includegraphics[width=\textwidth]{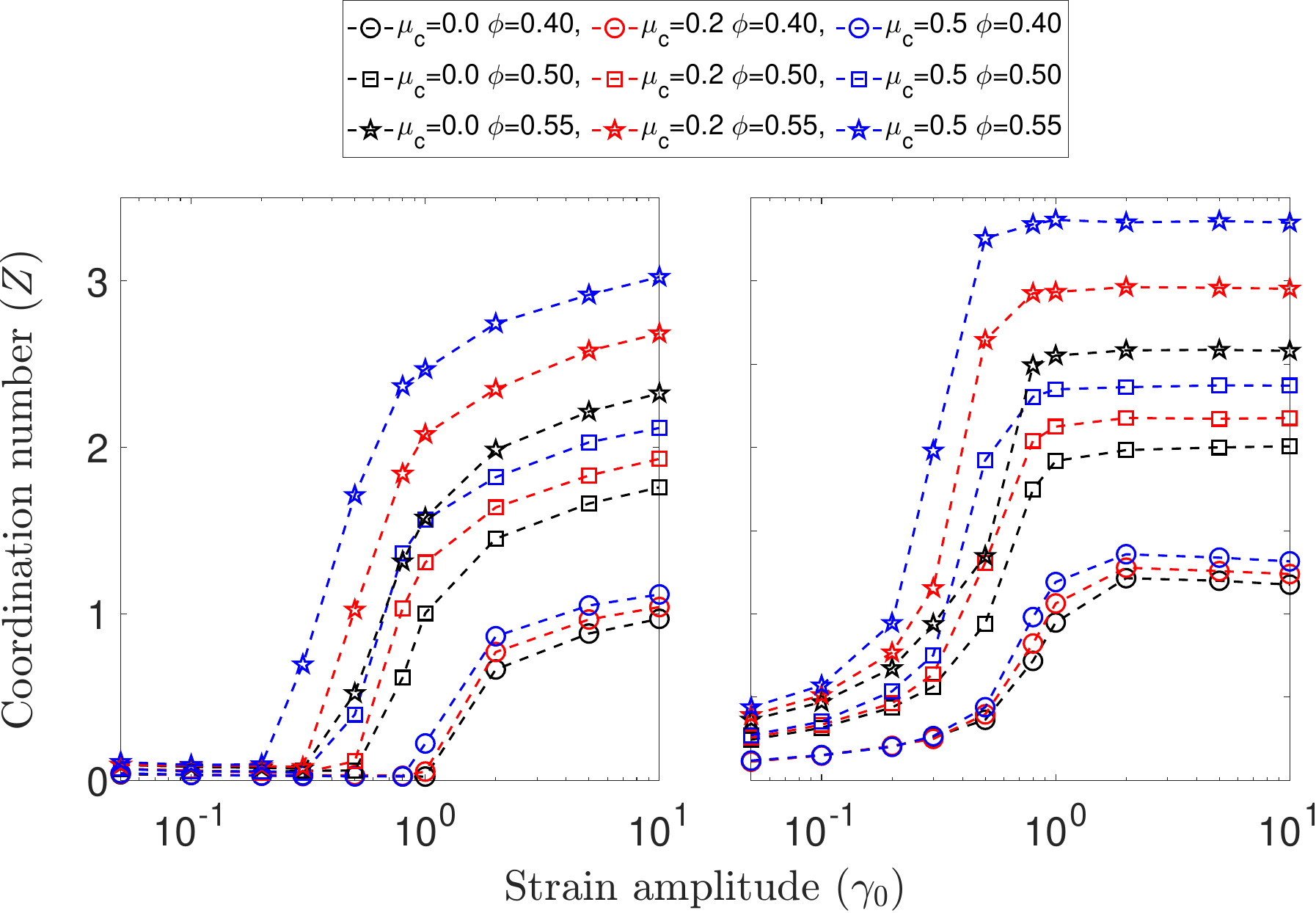}
  \caption{Coordination number (Z). Left: OS, right: RS. The coordination number is averaged over roughly $100$ strain units in the steady state.}  \label{f:contacts_cyc_amp_fric_phi_OS_RS}
\end{figure*}

The coordination number $Z$ as a function of $\gamma_{0}$ 
is displayed in figure \ref{f:contacts_cyc_amp_fric_phi_OS_RS}
for all the values of volume fraction and friction coefficient under investigation. For OS, $Z \approx 0$ at smaller $\gamma_0$, suggesting that the particles self-organize into contact-free absorbing states. In contrast, $Z>0$ even at low strain amplitudes in RS. 
This reveals why RIT is not observed in RS: in this case, we observe  non-hydrodynamic particle contacts, leading to diffusive dynamics, at all strain amplitudes. 
The particle contacts also affect the suspension viscosity. As shown in Figure \ref{f:stress_budget_OS_RS}, the contribution from the contact stress remains negligible for $\gamma_0<\gamma_{0,c}$ in OS, but it is non negligible for $\gamma_0<\gamma_{0,m}$ in RS. 
Since the total viscosity is mainly due to the hydrodynamic interactions at small amplitudes, the rheology is nevertheless similar in OS and RS.

\begin{figure}
  \centering
      \includegraphics[scale=0.8]{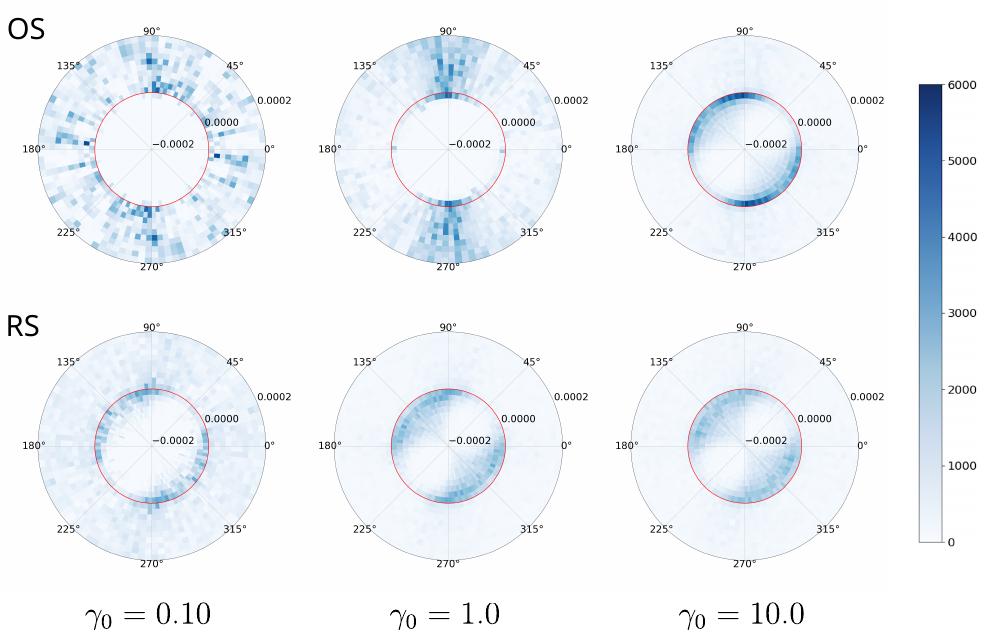} 
    \caption{Pairwise particle distribution $g(h,\theta)$. $\theta = 0^{\circ}$ and $90^{\circ}$ represent the flow and velocity gradient directions respectively. $\theta \in [0,\pi/2]$ or $[\pi, 3\pi/2]$ represents extensional quadrants and $\theta \in [\pi/2, \pi]$ or $[3\pi/2, 2\pi]$ represents the compressional qudrants. The contact point $h = 0$ is marked red, $h\leq0$ shows particle pairs in contact, and $h>0$ shows particles are separated. Figures are shown for suspension at $\phi=0.40, \mu_c=0.0$.}   
    \label{f:g_rad_ang_OS_RS}
\end{figure}

Next, we investigate the pairwise particle distribution $g(h,\theta)$ (see Figure \ref{f:g_rad_ang_OS_RS}), which provides information about the relative position of particle pairs in the suspension, projected onto the shear plane, as described in \S \ref{subsec:qoi}. 
The figure shows that particles are isotropically distributed and well-separated (thus the sparsity of data within a narrow gap) at the smallest strain amplitude under OS. 
Increasing the strain amplitude above $\gamma_{0,m}$ particles come into closer contact $(h\leq0)$ and form an anisotropic structure with an accumulation of particles around the compression axes ($\theta=3\pi/4$ and $7\pi/4$). These  observations are in line with the  previous findings for suspensions under simple shear \citep{morris2009review}.
Moreover,  we note that within the compression quadrants, neighboring particles tend to accumulate more towards $\theta\approx\pi/2$ and $3\pi/2$ than for $\theta \approx \pi$ or $\theta \approx 2\pi$, leading to more frequent collision along the velocity gradient direction than in the flow direction. As discussed by \citet{ge2022rheology} (see their Figure 3, inset), such anisotropic collision might explain the slightly positive first normal stress difference $N_1$ near $\gamma_{0,m}$ seen in Figure \ref{f:stress_diff_os}(a).

Although collisions  are enough to cause the suspension dynamics to transition from reversible to irreversible, our analysis provides a more detailed understanding of how the dynamics are related to the rheology as the strain amplitude is increased.
For RS, in contrast, the figure shows the development of  an anisotropic microstructure and particle contact even at the smaller strain amplitudes. 
Both the number of contacting pairs and the extent of anisotropy increase with $\gamma_0$, suggesting an analogue between RS and SS in terms of microstructal reorganization.

So far, we have examined and compared the rheology, dynamics and microstructre of OS and RS. These are the two extremes: the suspension undergoes sudden shear reversal in OS, whereas it undergoes gradual shear rotation and, ultimately, reversal in RS. Moreover, OS is a reversible protocol, while RS is not. To bridge the gap between the two, we therefore consider a combination of both, i.e.~reversible rotary shear (RRS), and present the main results in the following section.

\subsection{Reversible rotary shear (RRS)}
Reversible rotary shear (RRS), as already introduced in \S \ref{sec:model} is a combination of OS and RS, where the directions of both the flow and the shear rotation are reversed at half cycles. For comparison, we investigate the suspension under RRS protocol for $\phi=0.55$ and two friction coefficients $\mu_c = 0.0$ and $0.5$.

\begin{figure}
  \centering
  \begin{subfigure}{0.48\textwidth}
      \centering
      \includegraphics[width=\textwidth]{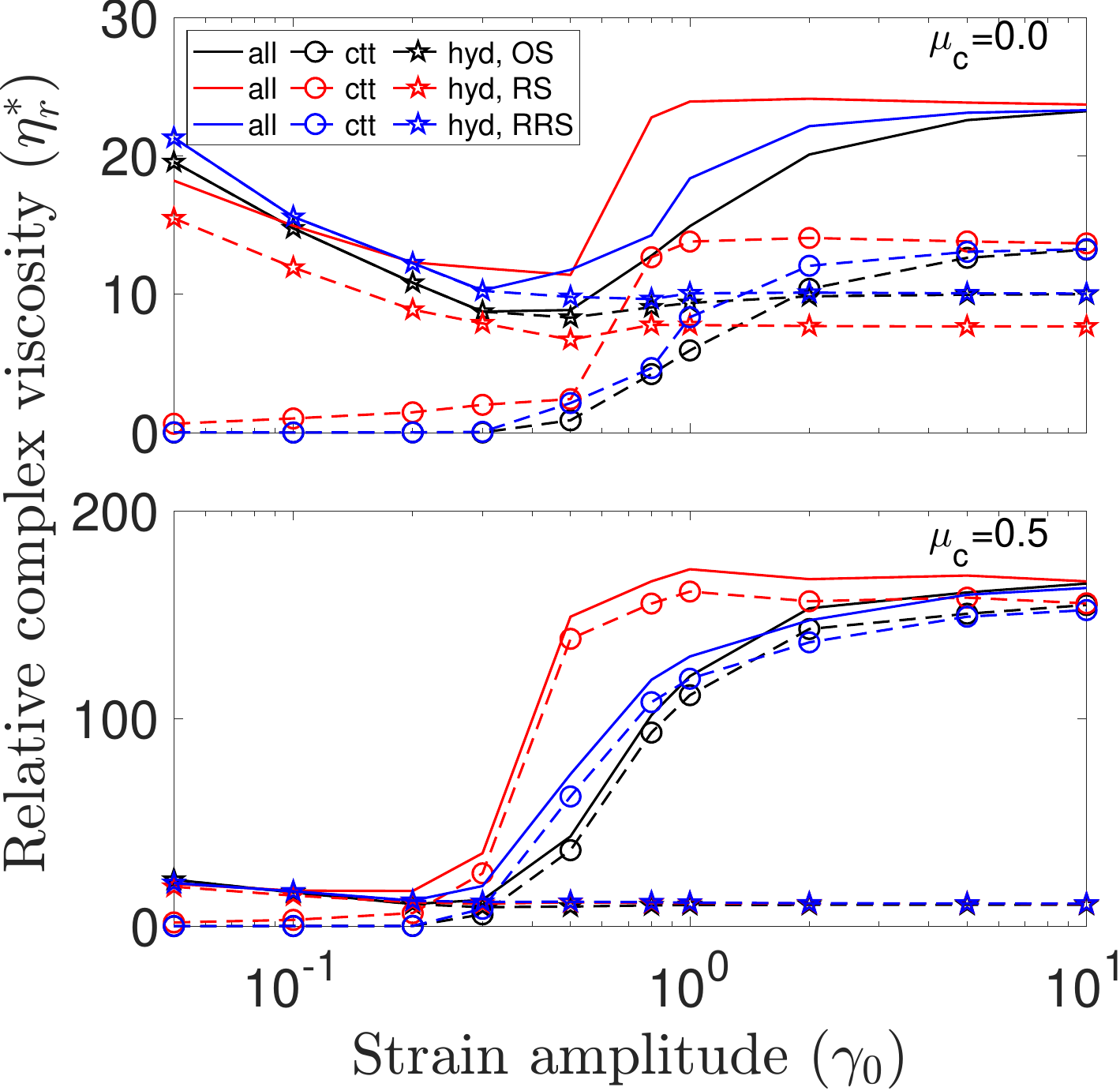} 
      \caption{}
      \label{f:stress_budget_OS_RS_RRS}
  \end{subfigure}
  \begin{subfigure}{0.48\textwidth}
      \centering
      \includegraphics[width=\textwidth]{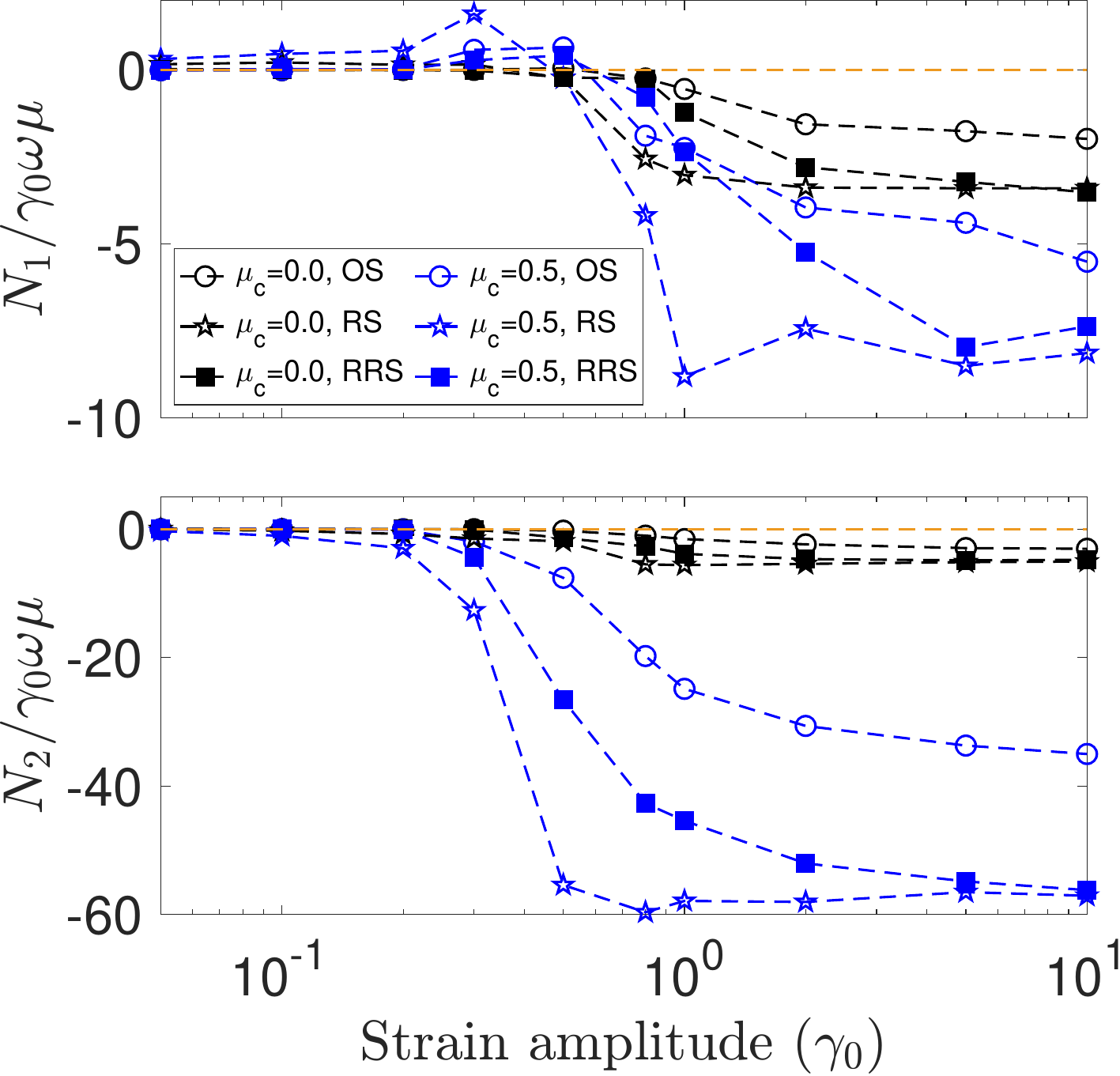} 
      \caption{}
      \label{f:stress_diff_OS_RS_RRS}
  \end{subfigure}  
  \caption{(a) Viscosity budget: total stresses (solid lines), contact stresses (circles), hydrodynamic stresses (stars) for OS (black lines), RS (red lines), and RRS (blue lines) at $\phi=0.55$. (b) Normal stress differences comparing OS, RS, and RRS for $\phi=0.55$ and $\mu_c=0.0, 0.5$: (top) first normal stress difference ($N_1$) and (bottom) second normal stress difference ($N_2$).}
  \label{f:rheology_OS_RS_RRS}
\end{figure}

As shown in Figure \ref{f:stress_budget_OS_RS_RRS}, suspensions undergoing RRS have a non-monotonic dependence of the complex viscosity  on the strain amplitute  similar to OS and RS: the complex viscosity decreases at smaller strain amplitudes, attains a minimum value at an intermediate strain amplitude and increases afterwards. As for the normal stress differences, we observe that both $N_1$ and $N_2$
are negative at large strain amplitudes, with $|N_1| \ll |N_2|$, as shown in Figure \ref{f:stress_diff_OS_RS_RRS}. From these observations we conclude that the suspension rheology is not sensitive to the different shearing protocols considered. 

\begin{figure}
    \centering
    \begin{subfigure}{0.48\textwidth}
      \centering
      \includegraphics[width=\textwidth]{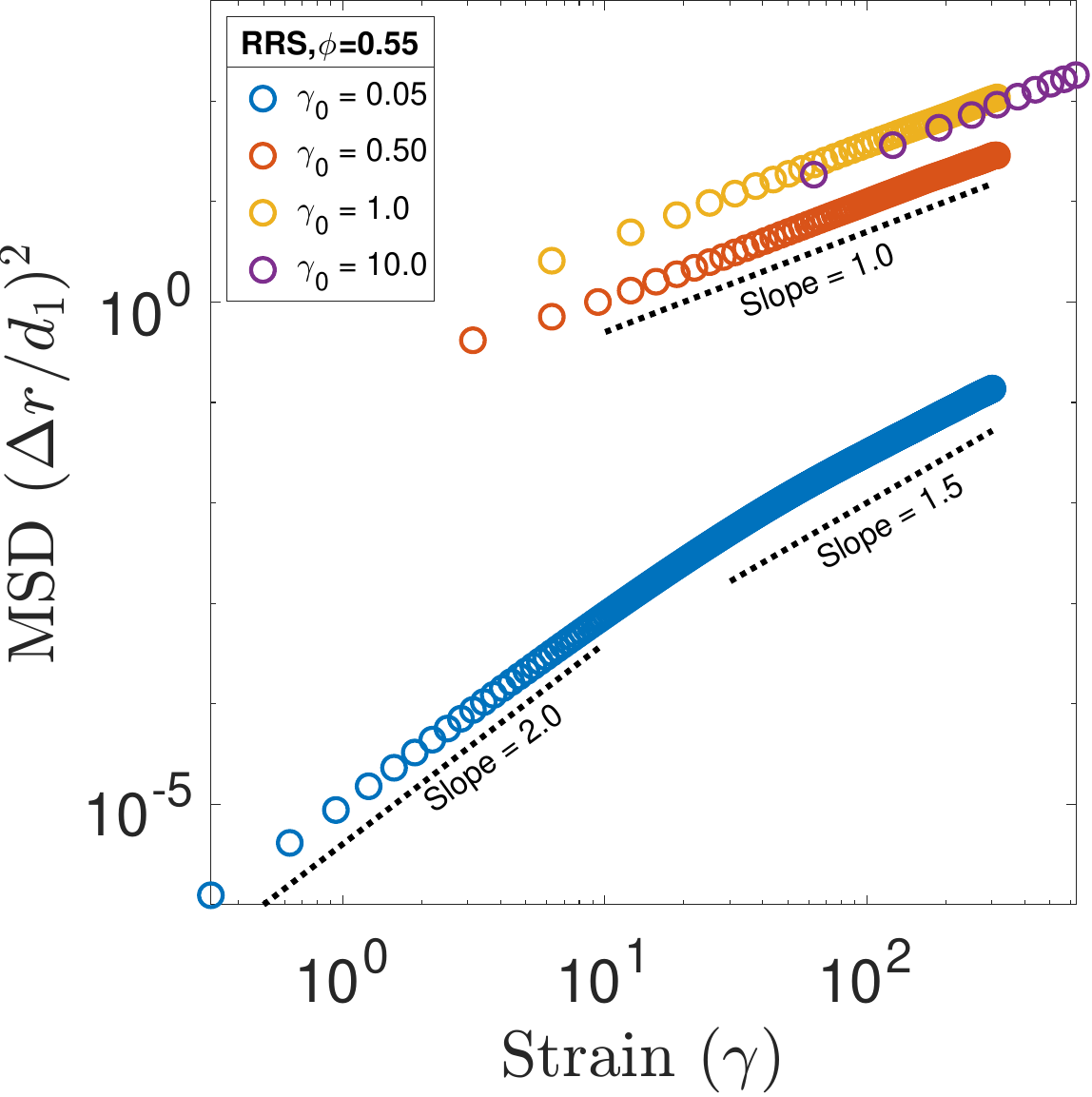} 
      \caption{}
      \label{subf:MSD_RRS}
    \end{subfigure}
    \hspace{0.2cm}
    \begin{subfigure}{0.48\textwidth}
      \centering
      \includegraphics[width=\textwidth]{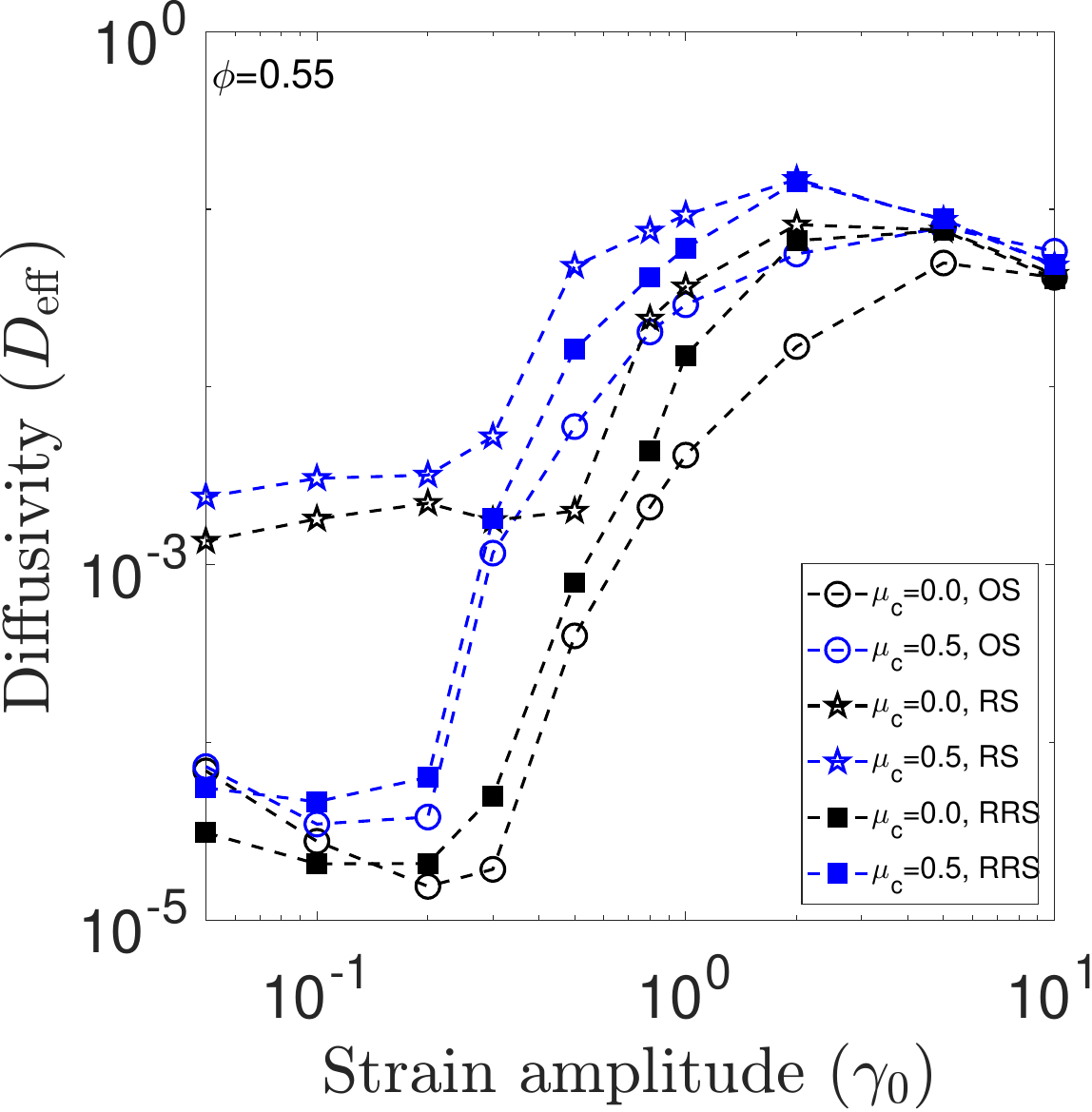} 
      \caption{}
      \label{subf:deff_rrs}
    \end{subfigure}
    \caption{(a) Particle mean-square displacement versus strain for suspensions undergoing reversible rotary shear (RRS) and the different values of the strain amplitude in the legend. (b) Comparison of the effective particle diffusivity for the three protocols under investigation: OS, RS and RRS.}
    \label{f:dynamics_rrs}
\end{figure}

Next, we investigate the dynamics of suspensions  undergoing RRS and compare them with the dynamics of the same suspension undergoing either OS or RS. The MSD versus strain are therefore shown in Figure \ref{subf:MSD_RRS}; the data reveal a similar trend as that observed for OS: the MSD does not scale linearly with the strain at smaller strain amplitudes, c.f.~the curve pertaining the case $\gamma_0 = 0.05$; it is however linear for larger strain amplitudes. Interestingly, the suspension under RRS experiences RIT as shown in Figure \ref{subf:deff_rrs}: the suspension is in reversible ``absorbing" state before a critical strain amplitude, $\gamma_0 < \gamma_{0,c}$, and in the irreversible ``diffusive" state at $\gamma_0 > \gamma_{0,c}$.  

\begin{figure}
\centering
  \includegraphics[width=\textwidth]{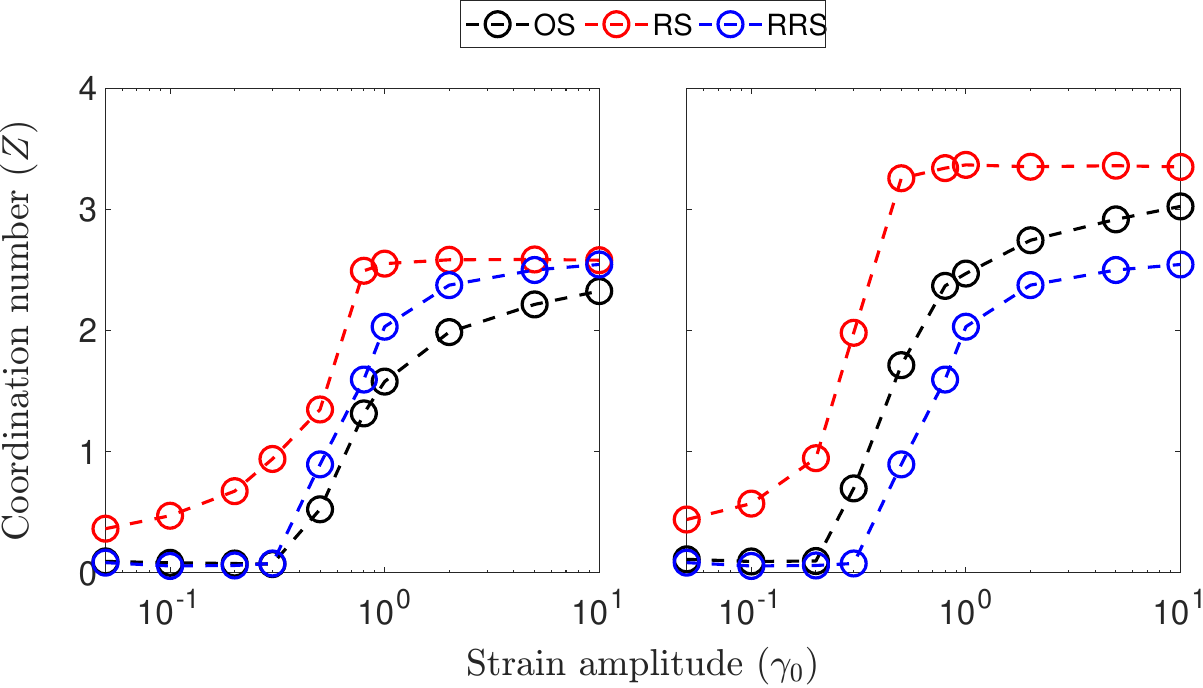}
\caption{Coordination number (Z) in OS, RS and RRS for $\phi=0.55$, $\mu_c=0.0,0.5$. The coordination number is averaged over roughly $100$ strain units in the steady state.} 
\label{f:contacts_amp}
\end{figure}

The above results show that suspensions driven by RRS can undergo the reversible-irreversible transition, as observed in OS but not in RS, despite they all share essentially the same rheology.
The different dynamical behavior of the suspension 
may be ascribed to the fundamental difference between the protocols: shear reversibility. The OS and the RRS have shear reversibility, whereas the RS does not. This suggests that absorbing states can only be reached via protocols that involve shear reversal. 
Hence, we can conclude that the difference that the shear reversal brings, is in the suspension microstructure. The coordination number $(Z)$ versus the strain amplitude $(\gamma_0)$ is shown in Figure \ref{f:contacts_amp} for the three protocols. The RRS protocol is, from all perspectives, similar to OS. At smaller strain amplitude the microstructure needs a long time to evolve but is broken before it develops by reversing the shear, which causes an immediate drop in the coordination number i.e. particle contacts; this mechanism is essential for the irreversible dynamics. When applying the RS protocol, the microstructure does not break, and the particle contacts grow continuously until saturation, similarly to the behavior in simple shear flows, which do not include shear reversal. This suggests that the differences between OS and RS are primarily due to the presence of sudden shear reversals in the former but not the latter, affecting both the dynamics at low strain amplitudes and the saturation to SS at higher strain amplitudes.

\section{Conclusion}\label{sec:conclusion}

In this work, we performed numerical simulations of dense, bidisperse suspensions of rigid, spherical particles at volume fractions ranging from 0.40 to 0.55 under three periodic flow protocols: the classical oscillatory shear (OS), the novel rotary shear (RS), and the reversible rotary shear (RRS), a combination of the first two.

Summarising the main findings, we observe the suspension rheology, both the viscosity and the normal stress differences, to be only weakly sensitive to the flow protocol, but strongly dependent on the applied strain amplitude. The viscosity shows a non-monotonic behavior with the strain amplitude, attaining a minimum viscosity at an intermediate strain amplitude ($\gamma_{0,m}$) that depends on the volume fraction. At the same strain amplitude $\gamma_{0,m}$, we observed the onset of second normal stress differences regardless of the shear protocol. However, the suspension dynamics change with the ﬂow protocol and the strain amplitude. Specifically, under OS and RRS, the suspensions show reversibility at low strain amplitudes and irreversibility at large strain amplitudes, thus undergoing a reversible-irreversible transition (RIT) at an intermediate strain amplitude ($\gamma_{0,c}$). In contrast, the suspension dynamics under the RS protocol are inherently irreversible at all strain amplitudes. The reason for this difference lies in the flow-induced suspension microstructure: for shear reversal protocols (OS and RRS), the particle pairs are more separated and isotropically distributed at smaller strain amplitudes, only to become anisotropic and in closer contact as the strain amplitude increases beyond a critical value. On the other hand, under the gradually varying RS protocol, particle pairs are anisotropically distributed and at contact at all strain amplitudes, even if the number of contacts is lower at low strain amplitudes. The presence of particle contacts at even the smallest strain amplitudes is the origin of the inherent irreversibility of RS, as also suggested in the literature based on simpler descriptions of particle interactions \citep{corte2008random, menon2009universality}.
Our RS and RSS protocols can be seen as a gentler form of shear rotations \citep{mari2014shear,acharya2024optimum}, resulting in a smoother behaviour than with sudden shear rotations.

For all protocols considered, we see the build-up of a large number of particle contacts and, hence, contact stress at strain amplitudes above a half. 
For OS, this strain amplitude is related to the strain needed to restructure contacts after a shear reversal. For the RS and RRS protocols, a corresponding flow direction reversal corresponds to half a rotation cycle, and interestingly, we see that a comparable accumulated strain over half a rotation cycle is needed to restructure the full contact network. This shows that the microstructure restructures on a comparable accumulated strain irrespectively of the shear modulations: gradual (as in our RS protocol) or sudden changes (as in shear reversals, \emph{e.g.,} OS). 

Finally, an important finding of the current work is that the presence of rheological characteristics such as minimum viscosity and the onset of second normal stress differences are not sufficient conditions for the suspensions to show dynamical characteristics such as RIT. To gain further understanding, the current work could therefore be extended to investigate the reversible rotary shear (RRS) protocol by varying the angles after which the shear is reversed. We expect that, after a certain number of rotations, the RRS would show dynamical features similar to RS.

\section*{Acknowledgments}
This project has received funding from the European Union’s Horizon 2020 research and innovation program under the Marie Skłodowska-Curie grant agreement No. 955605, YIELDGAP ITN network.
ZG is supported by an international postdoc grant from the Swedish Research Council (Grant No. 2021-06669VR).
MT acknowledges funding from the Swedish Research Council under grant No. 2021-04997. OT acknowledges funding from European Research Council through the  Starting Grant MUCUS (ERC-StG2019-852529).

\section*{Declaration of Interests} The authors report no conflict of interest

\section*{Data Availability Statement}

\appendix
\section{Forces}\label{app_sec:forces}

The individual force terms in the right-hand side of Eq. \ref{eq:force} acting on the $i^{th}$ particle of radius $a_i$, translational velocity ${\bf u}_i$, and angular velocity ${\bf w}_i$ are described here. The $j$ index is used for the interacting particle. For details, please refer to the previous studies \citep{seto2013discontinuous, mari2014shear,cheal2018rheology}.

\begin{enumerate}
    \item Stokes' drag:\\
    The Stokes drag force acts on each particle $i$ due to the surrounding fluid. It is given by,  
    \begin{equation}
    \begin{array}{ll}
        {\bf F}_i^S &= - 6 \pi \eta_0 a_i ({\bf u}_i - {\bf U}_\infty),\\[8pt]
    \end{array}
    \end{equation}
    
    where, $\eta_0$ is the dynamic viscosity of the fluid, ${\bf U}_\infty$ is the translational velocity of the underlying shear flow at the location of the particle centre of mass,  and $a_i$ is particle radius.   
    \item Lubrication force:\\
    The lubrication force acting on each lubricating particle pair $i,j$ is given as,  
    \begin{equation}
        \begin{array}{ll}
             {\bf F}^L_{ij} &= - (X_{ii}^A \mathbb{P}_n + Y_{ii}^A \mathbb{P}_t)({\bf u}_i - {\bf u}_j) + Y^B_{ii}({\bf w}_i \times {\bf n}_{ij}) + Y^B_{ji}({\bf w}_j \times {\bf n}_{ij}),\\[8pt]
            {\bf F^L_{ji}} &= -{\bf F^L_{ij}},\\[8pt]
        \end{array}
    \end{equation}

    where, ${\bf n}_{ij}$ denotes the unit normal vector pointing from particle $i$ to particle $j$, $\mathbb{P}_n = {\bf n}_{ij}{\bf n}_{ij}$ and $\mathbb{P}_t = \mathbb{I} - {\bf n}_{ij}{\bf n}_{ij}$ represent the normal and tangential projection matrices. The $X$'s and $Y$'s are scalar resistances depending on $\eta_0, a_i, a_j$ and the gap between the two particles, whose complete expressions are given in \citep{ge2020implementation}.
    
    \item Contact force:\\
    In case of small overlap between particles $i$ and $j$, equal and opposite contact forces are exerted on the pair, following Coulomb's friction law.

    \begin{subequations}
        \begin{equation}
            {\bf F}^C_{ij} = -k_n{\bf h}_{ij} - k_t \boldsymbol{\xi}_{ij},
        \end{equation}
        \begin{equation}
            {\bf F}^C_{ji} = -{\bf F}^C_{ij},
        \end{equation}
        \begin{equation}\label{eq:coulomb}
             |k_t \boldsymbol{\xi}_{ij}| \leq \mu_c |k_n{\bf h}_{ij}| ,
         \end{equation}
    \end{subequations}

In the expressions above, ${\bf h}_{ij} = h_{ij}{\bf n}_{ij}$ is the normal surface gap between particle i and j, $k_n$ the normal spring constant, $k_t$ the tangential spring constant and $\mu_c$ the friction coefficient.  The friction force ($k_t {\bf \xi}_{ij}$) comes as a consequence of particle roughness. Here we adopt the stick/slide model for friction force. The tangential stretch vector is calculated as
\begin{equation}
    \boldsymbol{ \xi}_{ij} = \left \{ \begin{array}{ll}
         \int_{t_0}^{t} -\mathbb{P}_t [(\textbf{u}_i-\textbf{u}_j)+(a_i \textbf{w}_i+a_j \textbf{w}_j) \times \textbf{n}_{ij}]dt', & \mbox{if} \  |\boldsymbol{ \xi}_{ij}| < |\boldsymbol{ \xi}_{max}|,\\
         \boldsymbol{ \xi}_{max} & \mbox{otherwise},
         \end{array} \right.
\end{equation}

where $\boldsymbol{\xi}_{max}$ comes from the Coulomb's law of friction in Eq.\ref{eq:coulomb}. \end{enumerate}

\section{Numerical implementations}\label{app_subsec:numericals}
The governing equations described in Eq. \ref{eq:force} are marched in time using the modified velocity-Verlet algorithm \citep{groot1997dissipative},

\begin{subequations}
    \begin{equation}
       \textbf{x}_i^{n+1} = \textbf{x}_i^{n} + \Delta t \textbf{u}_i^{n} + \frac{\Delta t^2}{2} \boldsymbol{ \alpha}_i^{n}
    \end{equation}
    \begin{equation}
        \textbf{u}_i^{n+\frac{1}{2}} = \textbf{u}_i^{n} + \frac{\Delta t}{2} \boldsymbol{\alpha}_i^{n}
    \end{equation}
    \begin{equation}
        \boldsymbol{\alpha}_i^{n+1} = \mathcal{F} \{\textbf{x}_i^{n+1},\textbf{u}_i^{n+\frac{1}{2}}\}
    \end{equation}
    \begin{equation}
        \textbf{u}_i^{n+1} = \textbf{u}_i^{n} + \frac{\Delta t}{2} (\boldsymbol{\alpha}_i^{n} + \boldsymbol{\alpha}_i^{n+1})
    \end{equation}
\end{subequations}

Here, $\textbf{x}_i,\textbf{u}_i, \boldsymbol{\alpha_i}$ denote the position, velocity and acceleration vector of the $i^{th}$ particle. The time steps are denoted by: $n, n+\frac{1}{2}, n+1$ are the current, half time step forward and one time step forward respectively, and $\mathcal{F}$ denote the force functional as in Eq. \ref{eq:force}. To simulate the shear flow and remove the wall effect we imposed Lees-Edwards \citep{lees1972computer} boundary condition on particle positions and velocity components as follows.

For OS,
\begin{subequations}\label{eq:os_periodicity}
    \begin{equation}\label{eq:os_y}
        y = \left \{
        \begin{array}{cc}
            mod\{(y + L_y - y'), L_y\},  &\mbox{if}\ z>L_z,\\[8pt] 
            mod\{(y + L_y + y'), L_y\},  &\mbox{if}\ z<0,\\[8pt]
            mod\{(y + L_y), L_y\},  &\mbox{otherwise},\\    
        \end{array} \right.              
    \end{equation}
    \begin{equation}\label{eq:os_z}
        z = mod\{(z + L_z), L_z\}, 
    \end{equation}
    \begin{equation}\label{eq:os_x}
        x = mod\{(x + L_x), L_x\}, 
    \end{equation}
    \begin{equation}\label{eq:os_v}
        u_y = \left \{
        \begin{array}{cc}
            u_y - u'_y, &\mbox{if}\ z>L_z,\\
            u_y + u'_y, &\mbox{if}\ z<0,\\
        \end{array}\right.
    \end{equation}
\end{subequations}

For RS and RRS, Eqs. (\ref{eq:os_y}), (\ref{eq:os_z}), (\ref{eq:os_v}) are used in addition to the shear periodicity in $x$ (secondary shear) direction,

\begin{subequations}\label{eq:rs_periodicity}
    \begin{equation}\label{eq:rs_x}
        x = \left \{
        \begin{array}{cc}
            mod\{(x + L_x - x'), L_x\},  &\mbox{if}\ z>L_z,\\[8pt] 
            mod\{(x + L_x + x'), L_x\},  &\mbox{if}\ z<0,\\[8pt]
            mod\{(x + L_x), L_x\},  &\mbox{otherwise},\\    
        \end{array} \right.               
    \end{equation}
    \begin{equation}\label{eq:rs_u}
        u_x = \left \{
        \begin{array}{cc}
            u_x - u'_x, &\mbox{if}\ z>L_z,\\[8pt]
            u_x + u'_x, &\mbox{if}\ z<0,\\
        \end{array}\right.
    \end{equation}
\end{subequations}

where $x'$ and $y'$ are the position shifts in the  $x$, and $y$ directions respectively, and $u'_x$ and $u_y'$ are the velocity shifts when crossing the periodic boundaries, given by
\begin{equation}
\begin{array}{llll}
    x' &= mod\{\dot{\gamma}_{xz} L_zt, L_x \}, &u_x' = \dot{\gamma}_{xz}L_z, \\[8pt] 
    y' &= mod\{\dot{\gamma}_{yz} L_zt, L_y \}, &u_y' = \dot{\gamma}_{yz}L_z. \\
\end{array}
\end{equation}

\section{Stress rotation}\label{app:stress_rot}
The stress tensor in continously rotating frame of flow-vorticity directions is given as:

\begin{equation}
    \sigma' = R \sigma R^T,
\end{equation}
where, the rotation matrix $R$ is given as:

\begin{equation}\label{eq:r_matrix}
    R = 
    \begin{bmatrix}
        \cos(-\theta) & \sin(-\theta) & 0 \\
        -\sin(-\theta) & \cos(-\theta) & 0 \\
        0 & 0 & 1 
    \end{bmatrix},
\end{equation}

where $\theta = \omega t$ in clockwise direction (therefore negative) for RS. Therefore, the stress tensor in the new basis is:

\begin{equation}\label{app_eq:stress_rot_1}
    \sigma' =  
        \begin{bmatrix}
            \sigma'_\text{vv} & \sigma'_\text{vf} & \sigma'_\text{vg}\\
            \sigma'_\text{fv} & \sigma'_\text{ff} & \sigma'_\text{fg}\\
            \sigma'_\text{gv} & \sigma'_\text{gf} & \sigma'_\text{gg}
        \end{bmatrix} \\
\end{equation}
\vspace{0.2in}
\begin{equation}\label{app_eq:stress_rot}
   \resizebox{\linewidth}{!}{$
    \sigma' = 
\begin{bmatrix}
\begin{array}{c} 
\sigma_{xx}\cos^2(\theta) + \sigma_{yy}\sin^2(\theta)\\ 
- (\sigma_{xy} + \sigma_{yx})\sin(\theta)\cos(\theta) 
\end{array} 
& 
\begin{array}{c} 
\sigma_{xx}\sin(\theta)\cos(\theta) - \sigma_{yx}\sin^2(\theta) \\ 
+ \sigma_{xy}\cos^2(\theta) - \sigma_{yy}\sin(\theta)\cos(\theta) 
\end{array} 
& 
\begin{array}{c} 
\sigma_{xz}\cos(\theta) \\ 
- \sigma_{yz}\sin(\theta) 
\end{array} 
\\[0.6cm]

\begin{array}{c} 
\sigma_{xx}\sin(\theta)\cos(\theta) + \sigma_{yx} \cos^2(\theta) \\ 
- \sigma_{xy}\sin^2(\theta) - \sigma_{yy}\sin(\theta)\cos(\theta)  
\end{array} 
& 
\begin{array}{c} 
\sigma_{xx}\sin^2(\theta) 
+ \sigma_{yy}\cos^2(\theta) 
\\ +(\sigma_{yx}+\sigma_{xy})\sin(\theta)\cos(\theta) 
\end{array} 
& 
\begin{array}{c} 
\sigma_{xz}\sin(\theta) \\ 
+ \sigma_{yz}\cos(\theta) 
\end{array} 
\\[0.6cm]

\begin{array}{c} 
\sigma_{zx}\cos(\theta) 
- \sigma_{zy}\sin(\theta) 
\end{array} 
& 
\begin{array}{c} 
\sigma_{zx}\sin(\theta) 
+ \sigma_{zy}\cos(\theta) 
\end{array} 
& 
\sigma_{zz} 
\end{bmatrix}
$}
\end{equation}

For RRS, $\theta$ takes the form:
\begin{equation}
\theta = \left \{
    \begin{array}{cc}
         \omega t & \mbox{if}\ 0 \leq t < T/2, \\
         \pi + \omega (T-t) & \mbox{if}\ T/2 \leq t < T, 
    \end{array}  
    \right.
\end{equation}

\section{Validations}\label{sec:validation}
To validate our simulations we compared the simulation results with that in literature. The rheology; relative complex viscosity ratio is compared with the experimental results by \cite{bricker2006oscillatory} and the dynamics; effective diffusivity is compared with the experimental and numerical results by \cite{pine2005chaos}.

\begin{figure}
\centering
\begin{subfigure}{.5\textwidth}
  \centering
  \includegraphics[width=\textwidth]{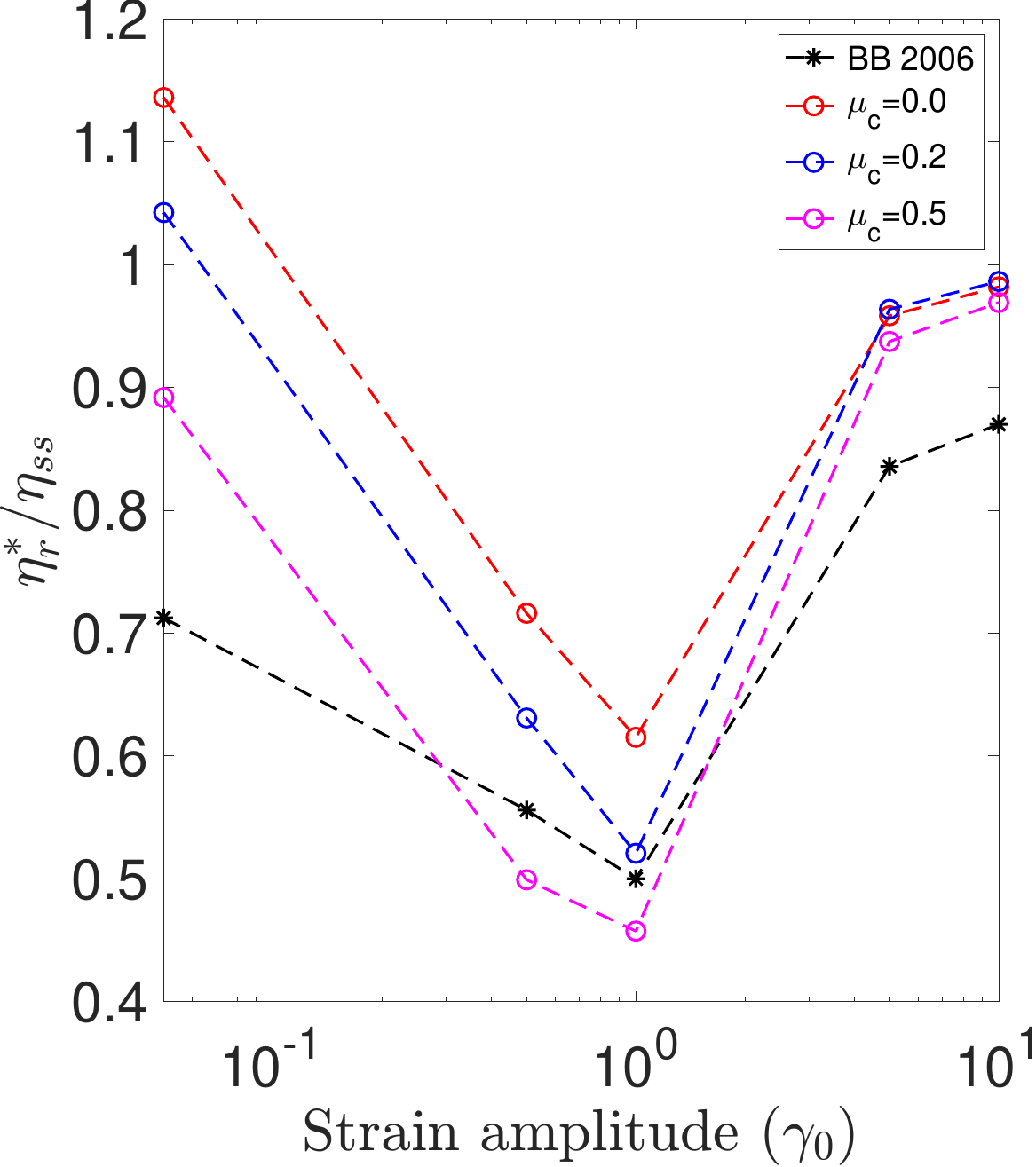}  
\caption{}
\label{subf:visc_BB}
\end{subfigure}
\vfill
\begin{subfigure}{0.49\textwidth}
  \centering
  \includegraphics[width=\textwidth]{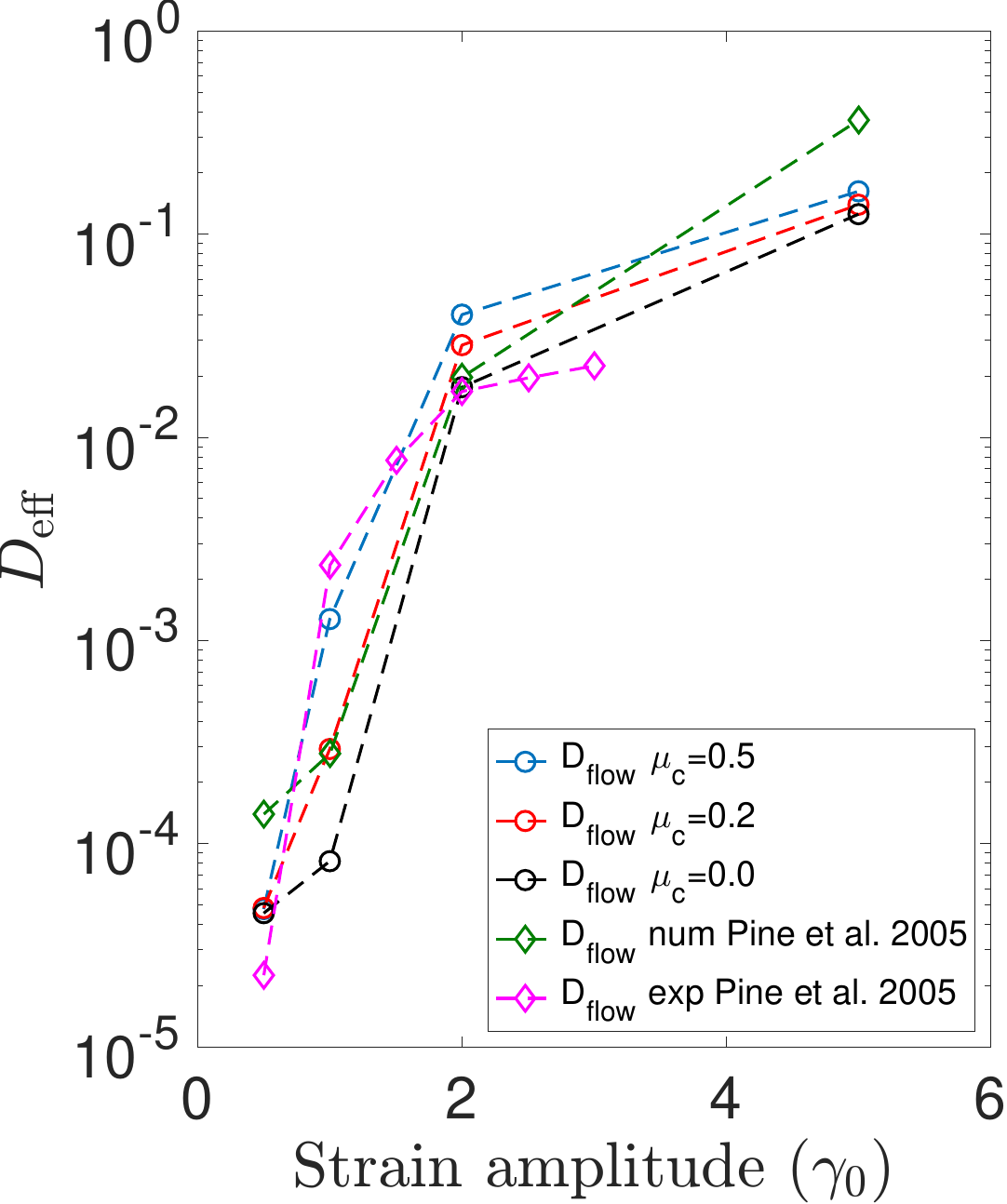} 
\caption{}
\label{subf:deff_flow}
\end{subfigure}
\begin{subfigure}{0.49\textwidth}
  \centering
  \includegraphics[width=\textwidth]{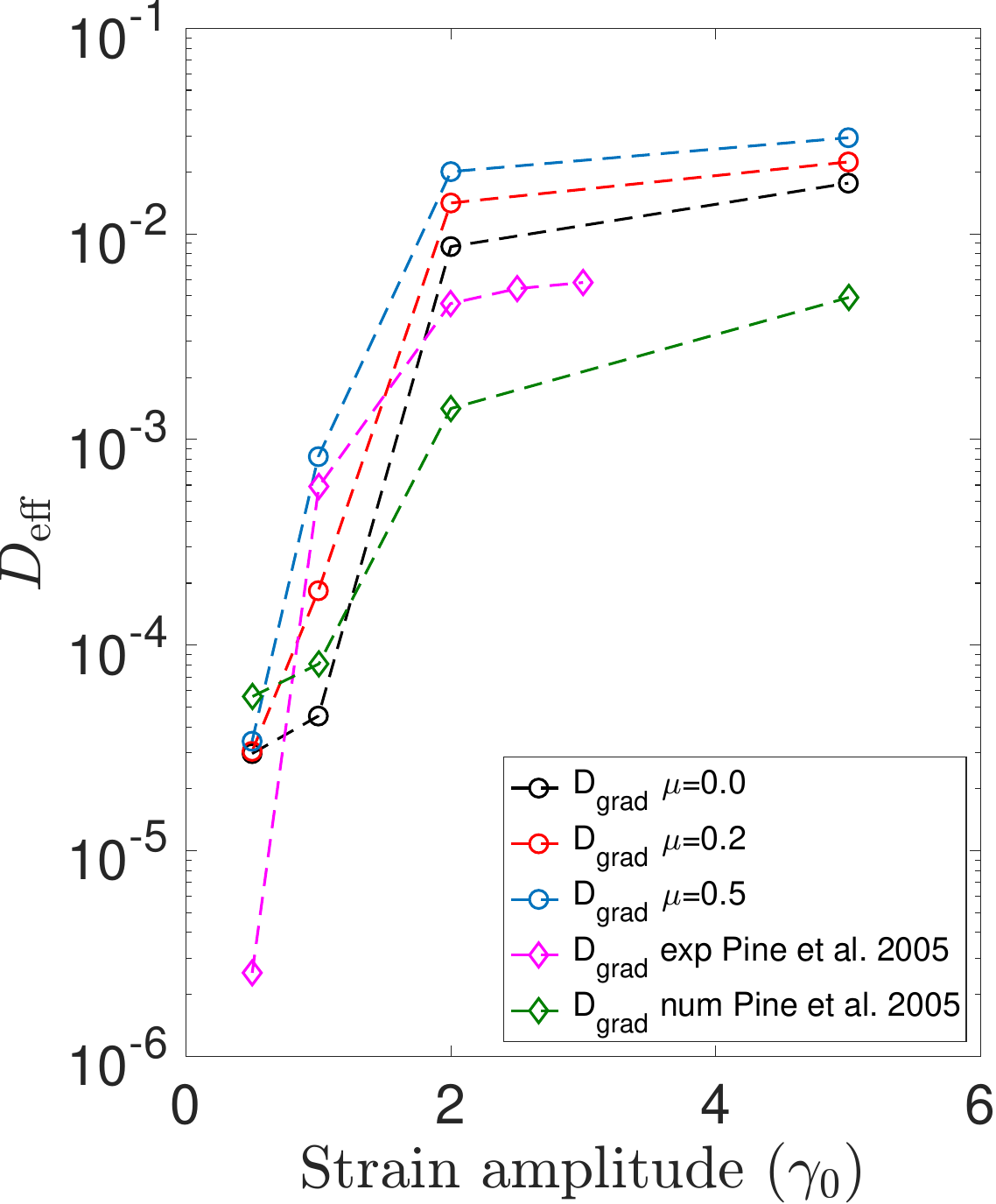} 
\caption{}
\label{subf:deff_grad}
\end{subfigure}
\caption{Validation at 40\% volume fraction in OS. (a) Complex viscosity ratio against the experiments by Bricker and Butler \citep{bricker2006oscillatory} and dynamics against experiments by Pine et al. \citep{pine2005chaos}, effective diffusion in (b) flow direction, (c) in the gradient direction for volume fraction $\phi=0.40$.}
\label{f:validations}
\end{figure}

Figure \ref{subf:visc_BB} compares the relative complex viscosity obtained from our simulations and experimental result from \citep{bricker2006oscillatory} for different strain amplitudes for volume fraction of $\phi=0.40$. The simulation results are shown for three friction coefficients $\mu_c = 0.0, 0.2$, and $0.5$. The complex viscosity of both works shows qualitative agreement. The relative complex viscosity shows non-monotonic behavior with respect to strain amplitude: it decreases at smaller $\gamma_0$, attains a minimum value at an intermediate $\gamma_0$, and then increases at the larger $\gamma_0$. The precise value of the viscosity not only depends upon $\gamma_0$ but on the physiochemical interactions between the particles as well.  

Figures \ref{subf:deff_flow} and \ref{subf:deff_grad} show effective diffusivity in flow and gradient directions respectively comparing our simulation results with the experimental and numerical results from \cite{pine2005chaos}. It can be seen that our simulation results closely match the numerical results from \citep{pine2005chaos}. However the quantitative difference could arise from the difference in shear protocols: they considered square wave shear protocol whereas we considered oscillatory shear protocol, the particle dispersity: we have bi-dispersed particles with a size ratio of $1.4$, whereas they considered a monodispersed system. However, their own numerical and experimental results have quantitative discrepancies owning to the machine precession, etc. Overall the rheology and dyanamics from our simulations closely follow the behavior observed in the literature. 

\begin{figure}
\begin{subfigure}{0.49\textwidth}
  \centering
  \includegraphics[width=\textwidth]{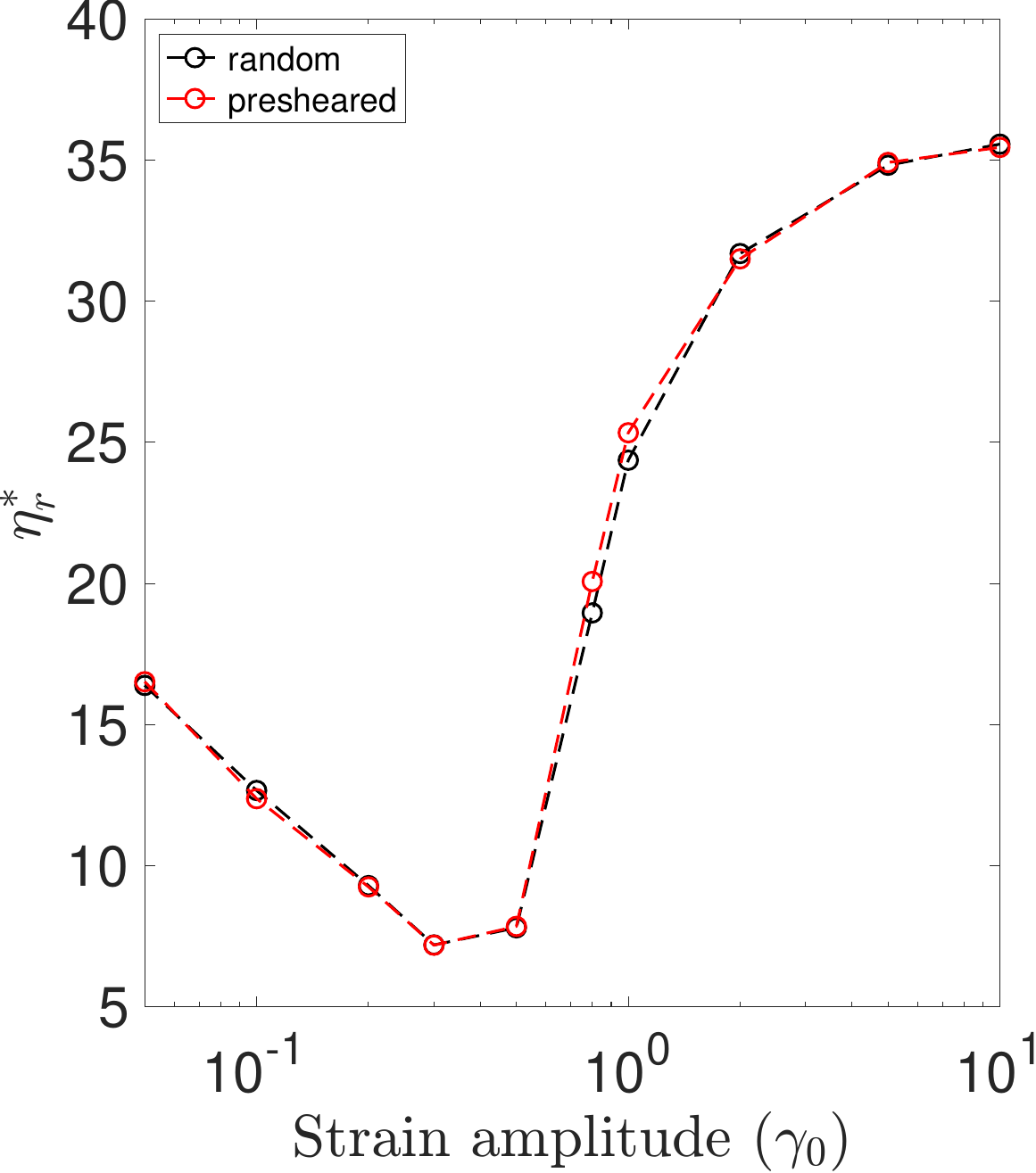}  
  \caption{}
\end{subfigure}
\begin{subfigure}{0.49\textwidth}
  \centering
  \includegraphics[width=\textwidth]{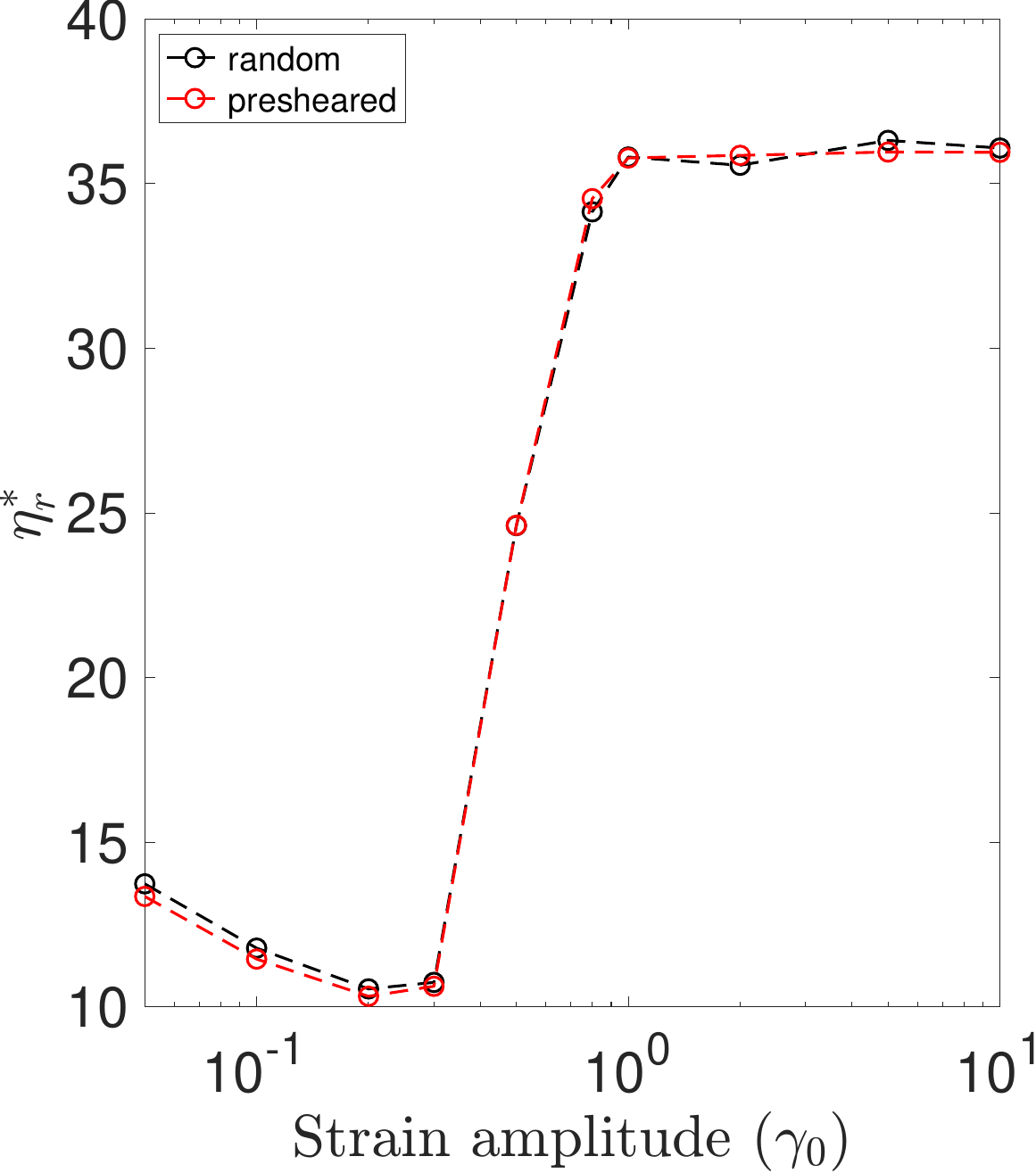}
  \caption{}
\end{subfigure}
\caption{Effect of initial condition on suspension rheology. Two initial conditions are compared for packing fraction ($\phi = 0.50$), and friction coefficient ($\mu_c=0.5$) for both (a) OS and (b) RS.} 
\label{f:effect_initial_conditions}
\end{figure}

We checked the initial condition dependency of mechanical response. Figure~\ref{f:effect_initial_conditions} shows the relative complex viscosity with respect to strain amplitude for two different initial conditions obtained by $(i)$ random distribution of particles and $(ii)$ after $40$ strain unit of pre-shearing for both OS and RS. For both OS and RS, we see that the steady-state results are independent of the initial conditions ~i.e. the suspension's mechanical response is independent of the particles' initial positions.  

\bibliographystyle{jfm}
\bibliography{jfm-instructions}

\end{document}